\documentclass[article]{aa}
\usepackage{graphicx}
\usepackage{longtable}
\begin{document}

   \title{Classical Cepheid Pulsation Models. XI. Effects of
   convection and chemical composition on the Period-Luminosity and Period-Wesenheit
   relations}

 \author{G. Fiorentino \inst{1,3}, M. Marconi \inst{2},
 I. Musella \inst{2}, F. Caputo \inst{3}}

\institute{INAF $-$ Osservatorio Astronomico di Bologna, Via
   Ranzani 1, 40127 Bologna, Italy;
   giuliana.fiorentino@oabo.inaf.it\\
\and INAF $-$ Osservatorio Astronomico Di Capodimonte, Via
Moiariello 16, 131 Napoli, Italy; marcella@na.astro.it;
ilaria@na.astro.it\\
\and INAF $-$ Osservatorio Astronomico di Roma, Via Frascati 33,
00040 Monte Porzio Catone, Italy;
caputo@mporzio.astro.it
 }

\date{}

\abstract{In spite of the relevance of Classical Cepheids as
primary distance indicators, a general consensus on the dependence
of the Period-Luminosity ($PL$) relation on the Cepheid chemical
composition has not been achieved yet.  From the theoretical point
of view, our previous investigations were able to reproduce some
empirical tests for suitable assumptions on the helium to metal
relative enrichment, but those results relied on specific
assumptions concerning the Mass-Luminosity relation and the
efficiency of the convective transfer in the pulsating envelopes.
%aims heading (mandatory)
In this paper, we investigate the effects of the assumed value of
the mixing length parameter $l/H_p$ on the pulsation properties
and we release the assumption of a fixed Mass-Luminosity relation.
%method
To this purpose, new nonlinear convective {\bf fundamental}
pulsation models have been computed for various chemical
compositions ($Z$=0.004, 0.008, 0.01 and 0.02) and adopting
$l/H_p$=1.7-1.8, which is larger than the one (1.5) used in our
previous papers.  From the extended model set, synthetic $PL$
relations in the various photometric bands are derived using the
predicted instability strip together with recent evolutionary
tracks.
%results heading (mandatory)
We show that as the $\l/H_p$ value increases the pulsation region
gets narrower, mostly due to the blueward shift of the red edge
for fundamental pulsation, with the effect becoming more important
at the higher metal contents ($Z\ge$ 0.01). However, the
comparison of the new models with the previously computed ones
shows that the $\l/H_p$ variation has no consequence on the
predicted Period-Wesenheit ($PW$) relations, which instead are
influenced by the pulsator metal content. On this ground, we
present a straightforward way to infer the distance and metal
content of variables with observed $BVI$ or $BVK$ magnitudes. As
for the $PL$ relations, we show that either the zero-point and the
slope are very slightly modified by the $\l/H_p$ variation, at
constant chemical composition. In the meanwhile, we confirm that:
(1) moving from visual to longer wavelengths, the predicted
Period-Magnitude distribution for a given metal content becomes
narrower and its slope becomes steeper; (2) decreasing the metal
content, the $PL$ relations become steeper and brighter, with the
amount of this metallicity effect decreasing from optical to
near-infrared bands.
%Conclusions
As a whole, {\bf we show that} our pulsation relations appear
fully consistent with the observed properties of Galactic and
Magellanic Cloud Cepheids, supporting the predicted steepening and
brightening of the $PL$ relations when moving from metal-rich to
metal-poor variables. Moreover, we show that the distances
inferred by the predicted $PW$ relations agree with recently
measured trigonometric parallaxes, whereas they suggest a
correction to the values based on the Infrared Surface Brightness
technique, as already found from an independent method. Finally,
also the pulsation metal contents suggested by the predicted $PW$
relations appear in statistical agreement with spectroscopic
[Fe/H] measurements. }

\authorrunning{Fiorentino et al.}
\titlerunning{Classical Cepheid Pulsation Models. XI}

\maketitle

%________________________________________________________________

\pagebreak
%\newpage

\section{Introduction}

Classical Cepheids are the most important distance indicators
within the Local Group as well as to external galaxies out to the
Virgo cluster. Moreover, {\bf they are used to calibrate a host of
secondary distance indicators, allowing us to reach cosmological
distances and to measure the Hubble constant}. However, Cepheid
distances are generally derived by adopting ``universal''
multiband Period-Luminosity ($PL$) relations calibrated on the
Large Magellanic Cloud (LMC) variables and even in the most recent
literature a general consensus on this crucial point has not been
achieved yet.

Regarding the slope of the $PL$ relations, for a long time it has
been assumed that it is constant over the total period range,
i.e., that the relations are linear. Conversely, recent studies
(Tammann \& Reindl 2002; Ngeow \& Kanbur 2004, 2005, 2006;
Sandage, Tammann \& Reindl 2004; Ngeow et al. 2005) show that in
the optical bands the LMC $PL$ relations are not linear, further
presenting evidence that the observed data are more consistent
with two linear relations and for a discontinuity at $P\sim$10
days. On the other hand, Persson et al. (2004, hereinafter P04)
show that the near-infrared ($JHK$) $PL$ relations of LMC Cepheids
with period from log$P\sim$0.4 to $\sim$2.0 are quite linear and
remarkably tight, leading Ngeow \& Kanbur (2006) to suggest that
the reason of such a behavior is because in the sample studied by
P04 there are not enough short-period variables.

Indeed, {\bf earlier discussions of the nonlinear convective
pulsation models computed by our group  (see Bono et al. 1999b
[Paper II]; Bono, Castellani \& Marconi 2000b [Paper III]; Caputo,
Marconi \& Musella 2000 [Paper V]) have already shown that the
instability strip boundaries in the log$L$-log$T_e$ plane are
almost linear, but when transformed in the various
Period-Magnitude planes they are better described by nonlinear
relations, mainly in the optical bands and at the red edge of
fundamental pulsation. Consequently, } we showed in Paper V that,
at fixed metal content: (1) the optical Period-Magnitude
distributions of the predicted pulsators are much better
represented by a quadratic line, (2) a discontinuity around
log$P\sim$ 1.4 should be adopted to constrain the theoretical
results into linear approximations, and (3) the predicted $PL$
relations become more and more linear and tight by moving from
optical to near-infrared bands. Those theoretical results have
been confirmed by more recent and complete pulsation models
(Fiorentino et al. 2002 [Paper VIII]; Marconi, Musella \&
Fiorentino 2005 [Paper IX]), including also the suggestion {\bf
that metal-poor ($Z$=0.004) Cepheids follow $PL$ relations which
are steeper and brighter than metal-rich ($Z$=0.02) ones, with the
amount of this metallicity effect decreasing from optical to
near-infrared bands.

From the observational side, several investigations (e.g., Tammann
et al. 2003; Sandage et al. 2004; Groenewegen et al. 2004; Ngeow
\& Kanbur 2004) point out the non-universality of the $PL$
relation, but with the Galactic Cepheid ($Z\sim$ 0.02) relations
steeper than the LMC ($Z\sim$ 0.008) counterparts. On the other
hand, Gieren et al. (2005, hereinafter G05), using the Infrared
Surface Brightness method to measure distances, affirm that the
$PL$ relation slope is the same both for Magellanic and Galactic
Cepheids. Moreover, a series of papers from the Araucaria Project
(see Pietrzynski et al. 2007 and references therein) devoted to
the survey of Cepheids in Local Group galaxies suggest no change
of the LMC $PL$ slope down to a Cepheid metal abundance of about
-1.0 dex. Regarding the zero-point, it has been suggested that
metal-rich Cepheids are somehow brighter, at fixed period, than
metal-poor variables (see Sasselov, Beaulieu \& Renault 1997;
Kennicutt et al. 1998, 2003; Kanbur et al. 2003; Storm et al. 2004
[S04]; Groenewegen et al. 2004; Sakai et al. 2004; Macri et al.
2006).  Although the comparison between theory and observations
will be discussed later, now we wish to} notice that the only two
{\it direct} tests of the metallicity effect are based on Cepheid
observations in the outer and inner fields in M101 (Kennicutt et
al. 1998) and NGC4258 (Macri et al. 2006) that show an oxygen
abundance varying from LMC-like values ([O/H]$\sim -$0.4, outer
fields) to sovrasolar values ([O/H]$\sim$ 0.2-0.3, inner fields),
which is just the range where our pulsation models suggest that
the sign of the metallicity dependence can be locally reversed. As
a fact, while early linear pulsation models (e.g. Sandage, Bell,
\& Tripicco 1999; Alibert et al. 1999; Baraffe \& Alibert 2001)
predict a moderate metallicity effect, the nonlinear convective
models computed by our group (see Paper V, Paper VIII and Paper IX
and references therein) show that the metallicity effect is not
linear over the range $Z$=0.004-0.04 but shows a ``turnover''
around $Z\sim$ 0.02. As discussed in the quoted papers, if the
LMC-calibrated $PL$ relations in the $V$ and $I$ bands are used to
infer the Cepheid distance, then the distance modulus of variables
with $Z<$ 0.008, which is the typical metal content of LMC
Cepheids, requires a {\it positive} correction whose amount
increases as the metallicity decreases, whereas for variables with
$Z\ge$ 0.02 the sign and amount of correction depend on the
adopted helium-to-metal enrichment ratio $\Delta Y/ \Delta Z$. As
a consequence, for very metal-rich ($Z\ge$ 0.03 or [O/H]$\ge$
+0.2) Cepheids the predicted metallicity correction to the
LMC-based distance modulus varies from $\sim-$ 0.15 mag to $\sim$
+0.25 mag as the ratio $\Delta Y/ \Delta Z$ increases from 2 to
3.5. On this ground, we showed that the empirical metallicity
correction suggested by Cepheid observations in the two M101
fields may be accounted for, provided that $\Delta Y/ \Delta
Z\sim$ 3.5. We add that recent spectroscopic [Fe/H] measurements
of Galactic Cepheids (Romaniello et al. 2005) seem to exclude that
the visual $P$-$M_V$ relation is independent of the metal content,
as well as that the metallicity correction follows the linear
relations suggested by the quoted empirical studies. Conversely, a
better agreement is found with the predicted non-linear correction
given in Paper VIII.

In spite of this promising agreement with observations, we cannot
ignore that the $PL$ relations predicted in our previous papers
are based on two fundamental assumptions: the pulsation models are
computed with a mixing-length parameter $\l/H_p$=1.5 and for each
mass the luminosity is fixed according to a canonical (i.e., no
mass-loss, no convective core overshooting) Mass-Luminosity ($ML$)
relation. We remind the reader that Classical Cepheids are post
Red Giant Branch (RGB) stars crossing the pulsation region of the
HR diagram during the characteristic ``blue loop'' connected with
the He-core burning. Since the luminosity of the loop mainly
depends on the original mass and chemical composition of the star,
the evolutionary models, as computed under specific assumptions,
yield a $ML$ relation which can be used to fix the luminosity of
pulsation models with a given mass and chemical composition.

Regarding the value of the mixing-length parameter, which is a
measure of the convection efficiency and is used to close the
system of equations describing the dynamical and convective
stellar structure, {\bf RR Lyrae pulsation models computed by our
group (see Caputo et al. 2000 [Paper V], Marconi et al. 2003; Di
Criscienzo, Marconi \& Caputo 2004a) show that variations of
$l/H_p$ lead to variations in the effective temperature of the
instability strip boundaries whose amount increase from the blue
to the red edge. Specifically, with $l/H_p$ in the range of 1.1 to
2.0 the effective temperatures at the fundamental blue edge (FBE)
and fundamental red edge (FRE) vary by
$\delta$log$T_e$(FBE)$\sim$0.02($l/H_p-$1.5) and
$\delta$log$T_e$(FRE)$\sim$0.04($l/H_p-$1.5), at constant
luminosity. Based on these results, we suspect that also the
Cepheid $PL$ relation, which is bound to the edges of the
instability strip, might depend on the adopted $l/H_p$ value.

On the other hand, specific modelling of the observed light curves
of both RR Lyrae stars and Classical Cepheids show that hotter
variables are well reproduced with $\l/H_p\sim$1.5, whereas
variables located toward the red part of the instability strip
require $\l/H_p\sim$1.8 (see Bono, Castellani \& Marconi 2000a,
2002; Castellani, Degl'innocenti, Marconi 2002; Di Criscienzo,
Marconi, Caputo 2004b; Marconi \& Clementini 2005; Natale et al.,
in preparation). These results are consistent with current
evolutionary models which adopt relatively high (1.9-2.2)
mixing-length values, as calibrated on observations. Thus}, in
order to investigate the effects of the adopted mixing-length
parameter, we have computed new fundamental pulsation models with
$Z$=0.004, 0.008, 0.01 and 0.02, $M$=5, 7, 9 and 11$M_{\odot}$,
and $\l/H_p$=1.7 and 1.8.

The comparison of these new models with those previously computed
at $\l/H_p$=1.5 is presented in Sect. 2. In this Section, we give
also the predicted relations involving the reddening insensitive
Wesenheit functions, while in Sect. 3 we present synthetic Cepheid
populations with selected chemical compositions and we discuss the
effects of the $\l/H_p$ parameter on the multiband $PL$ relations.
The comparison between theoretical results and observed Cepheids
is presented in Sect. 4, while the conclusions close the paper.

\section{The role of the mixing-length parameter}

The new fundamental pulsation models with $l/H_p$=1.7 and 1.8 are
listed in Table 1 together with $l/H_p$=1.5 computations presented
in previous papers. The input physics and computing procedures
have been already discussed (see Bono et al. 1999a [Paper I];
Paper III) and will not be repeated here. For each given chemical
composition and mass, as listed in the first three columns in the
Table, the adopted luminosity refers to the canonical $ML$
relation based on Bono et al. (2000c [B00]) and Castellani,
Chieffi \& Straniero (1992 [CCS]) evolutionary computations or
deals with higher luminosity levels (``over'') as produced by
convective core overshooting and/or mass loss (see later). The
entire pulsation region is covered by varying the model effective
temperature $T_e$ by steps of 100 K and, for the sake of the
following discussion, let us make clear that increasing
(decreasing) by 100 K the effective temperature of the computed
bluest (reddest) model yields non-pulsating structures.
Accordingly, we adopt the effective temperature of the computed
bluest model, increased by 50 K, as the effective temperature of
the fundamental blue edge (FBE), and the effective temperature of
the reddest model, decreased by 50 K, as the effective temperature
of the fundamental red edge (FRE). These values, which have the
intrinsic uncertainty of $\pm$ 50K, are listed in columns (7) and
(8) in Table 1.

\linespread{0.7}
\begin{table*} \label{modelli} \caption{Intrinsic parameters for
fundamental pulsating models.}
\begin{center}
%\vspace{0.5truecm}
\begin{tabular}{lcrcccccc}
\hline
\hline
\tiny{$Z$} & \tiny{$Y$} & \tiny{$M/M_{\odot}$} & \tiny{log$L/L_{\odot}$}&\tiny{$ML$}&\tiny{$l/H_p$} & \tiny{T$_{e,FRE}$ (K)} & \tiny{T$_{e,FBE}$ (K)}&  \tiny{References}\\
\hline
\tiny{0.004}& \tiny{  0.250   }&  \tiny{ 3.50}&   \tiny{ 2.61/2.86}& \tiny{ CCS/over}&   \tiny{1.5 }& \tiny{5800/5600}& \tiny{6000/6000}&   \tiny{M04}\\
\tiny{0.004}& \tiny{  0.250   }&  \tiny{ 4.00}&   \tiny{ 2.82    }&  \tiny{ CCS }&       \tiny{1.5 }& \tiny{5650     }& \tiny{5900    }&    \tiny{M04}\\
\tiny{0.004}&  \tiny{ 0.250    }&  \tiny{ 5.00}&   \tiny{ 3.07/3.30}& \tiny{ CCS/over}&   \tiny{1.5 }& \tiny{5600/5400}& \tiny{5900/6100 }& \tiny{Paper I  }\\
\tiny{0.004}& \tiny{  0.250   }&  \tiny{ 5.00}&   \tiny{ 3.24    }&  \tiny{ B00 }&       \tiny{1.8 }& \tiny{5500     }& \tiny{6000    }&    \tiny{This paper  }\\
\tiny{0.004}& \tiny{  0.250   }&  \tiny{ 7.00}&   \tiny{ 3.65/3.85}& \tiny{ CCS/over}&   \tiny{1.5 }& \tiny{5200/4900}& \tiny{5900/5900}&   \tiny{Paper I  }\\
\tiny{0.004}& \tiny{  0.250   }&  \tiny{ 7.00}&   \tiny{ 3.73    }&  \tiny{ B00 }&       \tiny{1.8 }& \tiny{5200     }& \tiny{5900    }&    \tiny{This paper  }\\
\tiny{0.004}&\tiny{   0.250   }& \tiny{  7.15}&  \tiny{ 3.73    }&  \tiny{ CCS }&       \tiny{1.5 }& \tiny{5100     }& \tiny{5900     }&    \tiny{M04 }\\
\tiny{0.004}&\tiny{   0.250   }& \tiny{  7.30}&  \tiny{ 3.76    }&  \tiny{ CCS  }&      \tiny{1.5 }& \tiny{5000     }& \tiny{5900     }&    \tiny{M04  }\\
\tiny{0.004}&\tiny{   0.250   }& \tiny{  7.45}&  \tiny{ 3.79    }&  \tiny{ CCS  }&      \tiny{1.5 }& \tiny{5000     }& \tiny{5900     }&    \tiny{M04  }\\
\tiny{0.004}&\tiny{   0.250   }& \tiny{  9.00}&  \tiny{ 4.00/4.25}& \tiny{ CCS/over}&   \tiny{1.5 }& \tiny{4900/4500}& \tiny{5800/5700 }&   \tiny{Paper I  }\\
\tiny{0.004}&\tiny{   0.250   }& \tiny{  9.00}&  \tiny{ 4.09    }&  \tiny{ B00 }&       \tiny{1.8 }& \tiny{4900}&      \tiny{5700}&         \tiny{This paper  }\\
\tiny{0.004}&\tiny{   0.250   }& \tiny{  11.00}& \tiny{ 4.40/4.65}& \tiny{ CCS/over}&   \tiny{1.5 }& \tiny{4400/4700}& \tiny{5500/5500 }&   \tiny{Paper I  }\\
\tiny{0.004}&\tiny{   0.250   }& \tiny{  11.00}& \tiny{ 4.39    }&  \tiny{ B00 }&       \tiny{1.8 }& \tiny{4600     }& \tiny{5500      }&   \tiny{This paper  }\\
\tiny{0.008}&\tiny{   0.250   }&\tiny{   3.50}& \tiny{  2.57/2.78}& \tiny{ CCS/over}&   \tiny{1.5 }& \tiny{5800/5600}& \tiny{5950/5900}&        \tiny{M04}\\
\tiny{0.008}&\tiny{   0.250   }& \tiny{  4.00 }& \tiny{ 2.78/3.07}& \tiny{ CCS/over }&  \tiny{1.5 }& \tiny{5650/5300 }&\tiny{5900/5900 }&   \tiny{M04}\\
\tiny{0.008}&\tiny{   0.250   }& \tiny{  5.00 }& \tiny{ 3.07/3.30}& \tiny{ CCS/over}&   \tiny{1.5}&  \tiny{5500/5300 }&\tiny{5900/6000 }&   \tiny{Paper I  }\\
\tiny{0.008}&\tiny{   0.250   }& \tiny{  5.00 }& \tiny{ 3.14    }&  \tiny{ B00  }&      \tiny{1.8 }& \tiny{5500 }&     \tiny{5900 }&        \tiny{This paper  }\\
\tiny{0.008}&\tiny{   0.250   }& \tiny{  6.55 }& \tiny{ 3.55    }&\tiny{ CCS  }&        \tiny{1.5 }& \tiny{5100 }&     \tiny{5900  }&       \tiny{ Paper VI  }\\
\tiny{0.008}&\tiny{   0.250   }& \tiny{  6.70 }& \tiny{ 3.59    }&  \tiny{ CCS  }&      \tiny{1.5 }& \tiny{5100 }&     \tiny{5900  }&       \tiny{ Paper VI }\\
\tiny{0.008}&\tiny{   0.250   }& \tiny{  6.85 }& \tiny{ 3.62    }&  \tiny{ CCS  }&      \tiny{1.5 }& \tiny{5100 }&     \tiny{5800  }&       \tiny{ Paper VI  }\\
\tiny{0.008}&\tiny{   0.250   }& \tiny{  7.00 }& \tiny{ 3.65/3.85}& \tiny{ CCS/over}&   \tiny{1.5 }& \tiny{5000/4700 }&\tiny{5800/5700 }&   \tiny{ Paper I  }\\
\tiny{0.008}&\tiny{   0.250   }& \tiny{  7.00 }& \tiny{ 3.62    }&  \tiny{ B00  }&      \tiny{1.8 }& \tiny{5200 }&     \tiny{5800 }&        \tiny{ This paper  }\\
\tiny{0.008}&\tiny{   0.250   }& \tiny{  7.15 }& \tiny{ 3.69    }&  \tiny{ CCS }&       \tiny{1.5 }& \tiny{4900}&      \tiny{5800 }&        \tiny{ Paper VI  }\\
\tiny{0.008}&\tiny{   0.250   }& \tiny{  7.30 }& \tiny{ 3.72    }&  \tiny{ CCS }&       \tiny{1.5 }& \tiny{4900}&      \tiny{5800  }&       \tiny{ Paper VI  }\\
\tiny{0.008}&\tiny{   0.250   }& \tiny{  7.45 }& \tiny{ 3.75    }&  \tiny{ CCS }&       \tiny{1.5 }& \tiny{4900}&      \tiny{5800  }&       \tiny{ Paper VI  }\\
\tiny{0.008}&\tiny{   0.250   }& \tiny{  9.00 }& \tiny{ 4.00/4.25}&  \tiny{ CCS/over}&   \tiny{1.5 }& \tiny{4700/4200}& \tiny{5600/5300 }&   \tiny{  Paper I  }\\
\tiny{0.008}&\tiny{   0.250   }& \tiny{  9.00 }& \tiny{ 3.99    }&  \tiny{ B00 }&       \tiny{1.8 }& \tiny{4900 }&     \tiny{5600 }&        \tiny{  This paper  }\\
\tiny{0.008}&\tiny{   0.250   }& \tiny{  11.00}& \tiny{ 4.40/4.65}&  \tiny{ CCS/over }&  \tiny{1.5 }& \tiny{4600/4200}& \tiny{5200/5200}&    \tiny{ Paper I  }\\
\tiny{0.008}&\tiny{   0.250   }& \tiny{  11.00}& \tiny{ 4.28    }&  \tiny{ B00 }&       \tiny{1.8 }& \tiny{4600 }&     \tiny{5400}&         \tiny{ This paper  }\\
\tiny{0.01} &\tiny{   0.260     }& \tiny{  5.00 }& \tiny{ 3.13    }&  \tiny{ B00 }&       \tiny{1.5 }& \tiny{5300}&      \tiny{5800 }&      \tiny{ Paper IX  }\\
\tiny{0.01} &\tiny{   0.260     }& \tiny{  7.00 }& \tiny{ 3.61    }&  \tiny{ B00 }&       \tiny{1.5 }& \tiny{4900}&      \tiny{5900}&       \tiny{ Paper IX  }\\
\tiny{0.01} &\tiny{   0.260     }& \tiny{  9.00 }& \tiny{ 3.98    }&  \tiny{ B00 }&       \tiny{1.5 }& \tiny{4500 }&     \tiny{5600 }&      \tiny{ Paper IX }\\
\tiny{0.01} &\tiny{   0.260     }& \tiny{  11.00}& \tiny{ 4.27    }&  \tiny{ B00 }&       \tiny{1.5 }& \tiny{4200}&      \tiny{5400}&       \tiny{ Paper IX }\\
\tiny{0.01} &\tiny{   0.260     }& \tiny{  5.00 }& \tiny{ 3.13    }&  \tiny{ B00 }&       \tiny{1.8 }& \tiny{5500}&      \tiny{5900}&       \tiny{ This paper  }\\
\tiny{0.01} &\tiny{   0.260     }& \tiny{  7.00 }& \tiny{ 3.61    }&  \tiny{ B00 }&       \tiny{1.8 }& \tiny{5200}&      \tiny{5800}&       \tiny{ This paper  }\\
\tiny{0.01}& \tiny{  0.260     }&  \tiny{ 9.00 }&  \tiny{ 3.98    }&  \tiny{ B00 }&       \tiny{1.8 }& \tiny{4900 }&     \tiny{5400}&       \tiny{ This paper  }\\
\tiny{0.01}& \tiny{  0.260     }&  \tiny{ 11.00}&  \tiny{ 4.27    }&  \tiny{ B00 }&       \tiny{1.8 }& \tiny{4500}&      \tiny{5100}&       \tiny{ This paper  }\\
\tiny{0.02}& \tiny{  0.250     }&  \tiny{ 5.00 }&  \tiny{ 3.00    }&  \tiny{ B00 }&       \tiny{1.5 }& \tiny{5300}&      \tiny{5700 }&      \tiny{ Paper IX        }\\
\tiny{0.02}& \tiny{  0.250     }&  \tiny{ 7.00 }&  \tiny{ 3.49    }&  \tiny{ B00 }&       \tiny{1.5 }& \tiny{4800}&      \tiny{5600 }&      \tiny{ Paper IX       }\\
\tiny{0.02}& \tiny{  0.250     }&  \tiny{ 9.00 }&  \tiny{ 3.86    }&  \tiny{ B00 }&       \tiny{1.5 }& \tiny{4400}&      \tiny{5300 }&      \tiny{ Paper IX       }\\
\tiny{0.02}& \tiny{  0.250     }&  \tiny{ 11.00}&  \tiny{ 4.15    }&  \tiny{ B00 }&       \tiny{1.5 }& \tiny{4300}&      \tiny{5000 }&      \tiny{ Paper IX      }\\
\tiny{0.02}& \tiny{  0.260     }&  \tiny{ 5.00 }&  \tiny{ 3.02    }&  \tiny{ B00 }&       \tiny{1.5 }& \tiny{5400}&      \tiny{5800  }&     \tiny{ Paper IX       }\\
\tiny{0.02}& \tiny{  0.260     }&  \tiny{ 7.00 }&  \tiny{ 3.51    }&  \tiny{ B00 }&       \tiny{1.5 }& \tiny{4900}&      \tiny{5600  }&     \tiny{ Paper IX       }\\
\tiny{0.02}& \tiny{  0.260     }&  \tiny{ 9.00 }&  \tiny{ 3.88    }&  \tiny{ B00 }&       \tiny{1.5 }& \tiny{4600}&      \tiny{5500  }&     \tiny{ Paper IX       }\\
\tiny{0.02}& \tiny{  0.260     }&  \tiny{ 11.00}&  \tiny{ 4.17    }&  \tiny{ B00 }&       \tiny{1.5 }& \tiny{4300}&      \tiny{5300  }&     \tiny{ Paper IX       }\\
\tiny{0.02}& \tiny{  0.280     }&  \tiny{ 4.00 }&  \tiny{ 2.97    }&  \tiny{ over}&       \tiny{1.5 }& \tiny{5300}&      \tiny{5800 }&      \tiny{ B01       }\\
\tiny{0.02}& \tiny{  0.280     }&  \tiny{ 4.50 }&  \tiny{ 2.90    }&  \tiny{ CCS }&       \tiny{1.5 }& \tiny{5500}&      \tiny{5800 }&      \tiny{ B01     }\\
\tiny{0.02}& \tiny{  0.280     }&  \tiny{ 5.00 }&  \tiny{ 3.07/3.30}& \tiny{ CCS/over }&  \tiny{1.5 }& \tiny{5400/5400}& \tiny{5900/5700 }& \tiny{ Paper I        }\\
\tiny{0.02}& \tiny{  0.280     }&  \tiny{ 5.00 }&  \tiny{ 3.07    }&  \tiny{ B00 }&       \tiny{1.7 }& \tiny{5400}&      \tiny{5900 }&      \tiny{ This paper    }\\
\tiny{0.02}& \tiny{  0.280     }&  \tiny{ 5.00 }&  \tiny{ 3.07    }&  \tiny{ B00 }&       \tiny{1.8 }& \tiny{5550}&      \tiny{5950 }&      \tiny{ This paper    }\\
\tiny{0.02}& \tiny{  0.280     }&  \tiny{ 6.25 }&  \tiny{ 3.42    }&  \tiny{ CCS }&       \tiny{1.5 }& \tiny{5000}&      \tiny{5500 }&      \tiny{ B01     }\\
\tiny{0.02}& \tiny{  0.280     }&  \tiny{ 6.50 }&  \tiny{ 3.48    }&  \tiny{ CCS }&       \tiny{1.5 }& \tiny{5000}&      \tiny{5600 }&      \tiny{ B01      }\\
\tiny{0.02}& \tiny{  0.280     }&  \tiny{ 6.75 }&  \tiny{ 3.54    }&  \tiny{ CCS }&       \tiny{1.5 }& \tiny{4900}&      \tiny{5550 }&      \tiny{ B01      }\\
\tiny{0.02}& \tiny{  0.280     }&  \tiny{ 7.00 }&  \tiny{ 3.65/3.85}& \tiny{ CCS/over }&  \tiny{1.5 }& \tiny{4700/4400}& \tiny{5400/5200 }& \tiny{ Paper I    }\\
\tiny{0.02}& \tiny{  0.280     }&  \tiny{ 7.00 }&  \tiny{ 3.56    }&  \tiny{ B00 }&       \tiny{1.7 }& \tiny{5000}&      \tiny{5400}&       \tiny{ This paper    }\\
\tiny{0.02}& \tiny{  0.280     }&  \tiny{ 7.00 }&  \tiny{ 3.56    }&  \tiny{ B00 }&       \tiny{1.8 }& \tiny{5150}&      \tiny{5350}&       \tiny{ This paper    }\\
\tiny{0.02}& \tiny{  0.280     }&  \tiny{ 9.00 }&  \tiny{ 4.00/4.25}& \tiny{ CCS/over }&  \tiny{1.5 }& \tiny{4400/3900}& \tiny{5100/4900  }&\tiny{ Paper I       }\\
\tiny{0.02}& \tiny{  0.280     }&  \tiny{ 9.00 }&  \tiny{ 3.92    }&  \tiny{ B00 }&       \tiny{1.7 }& \tiny{4700}&      \tiny{5000 }&      \tiny{ This paper    }\\
\tiny{0.02}& \tiny{  0.280     }&  \tiny{ 11.00}&  \tiny{ 4.40   }&   \tiny{ CCS  }&      \tiny{1.5 }& \tiny{3900}&      \tiny{4800}&       \tiny{ Paper I        }\\
\tiny{0.02}& \tiny{  0.280       }&  \tiny{ 11.00}&  \tiny{ 4.21    }&  \tiny{ B00 }&       \tiny{1.7 }& \tiny{4300}&      \tiny{4700}&     \tiny{ This paper    }\\
\tiny{0.02}& \tiny{  0.310       }&  \tiny{ 5.00 }&  \tiny{ 3.13    }&  \tiny{ B00 }&       \tiny{1.5 }& \tiny{5400  }&    \tiny{5900  }&   \tiny{ Paper VIII  }\\
\tiny{0.02}& \tiny{  0.310       }&  \tiny{ 7.00 }&  \tiny{ 3.62    }&  \tiny{ B00 }&       \tiny{1.5 }& \tiny{4900  }&    \tiny{5600  }&   \tiny{ Paper VIII }\\
\tiny{0.02}& \tiny{  0.310       }&  \tiny{ 9.00 }&  \tiny{ 3.98    }&  \tiny{ B00 }&       \tiny{1.5 }& \tiny{4500  }&    \tiny{5300  }&   \tiny{ Paper VIII  }\\
\tiny{0.02}& \tiny{  0.310       }&  \tiny{ 11.00}&  \tiny{ 4.27    }&  \tiny{ B00 }&       \tiny{1.5 }& \tiny{4200  }&    \tiny{5100  }&   \tiny{ Paper VIII }\\
\hline
\end{tabular}
\end{center}
Note$-$ B01: Bono et al. (2001); Paper VI: Bono, Marconi \&
Stellingwerf (2000d); M04: Marconi et al. (2004).
%\end{flushleft}
\end{table*}
\linespread{1}

%fig1
\begin{figure}
\includegraphics[width=8cm]{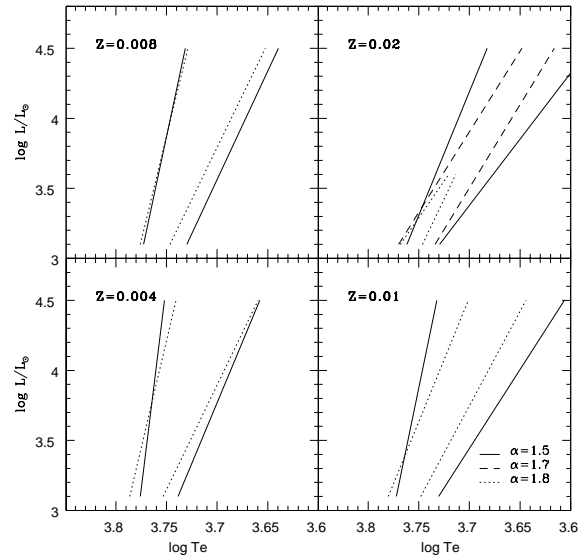}
\caption{\small{HR diagram of the fundamental pulsation region
when varying the value of the $l/H_p$ parameter as labelled, at
four different  chemical compositions. The plotted lines are
linear interpolations of the computed boundaries.}}
\label{7587fig1}
\end{figure}

\subsection{Instability strip and light curves}

Figure 1 shows the HR diagram of the fundamental pulsation region
for different assumptions on the $\l/H_p$ parameter, at constant
chemical composition. The lines depicting the edges of the
instability strip are drawn by linear interpolation. We find that
the effects of the $\l/H_p$ value increased from 1.5 to 1.8 are
negligible at $Z$= 0.004, whereas they become significant when
moving to larger metal contents. With $Z$=0.01 the FRE   gets bluer
by $\sim$300 K and the FBE redder by $\sim$100 K, while at $Z$=0.02
the effects are even more important. As a fact, yet with
$\l/H_p$=1.7 the FBE becomes redder by $\sim$100 K and the FRE bluer
by $\sim$300 K. Note that a further increase to $\l/H_p$=1.8 yields
that the pulsation is completely inhibited for masses larger than
7$M_{\odot}$. This result is due to the higher opacity of metal-rich
stellar envelopes that, in turn, produces an higher efficiency of
convection in damping pulsation.

%fig2
\begin{figure}
\includegraphics[width=9cm]{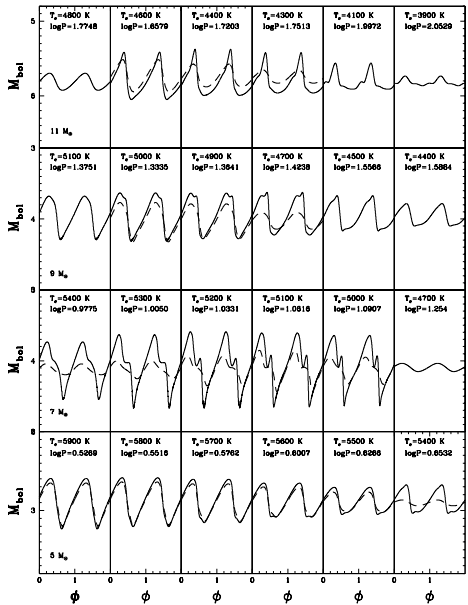}
\caption{\small{Effects of varying $l/H_p$ from 1.5 (solid lines) to
1.7 (dashed lines) on fundamental bolometric light curves at
$Z$=0.02.}} \label{7587fig2}
\end{figure}

%fig3
\begin{figure}
\includegraphics[width=9cm]{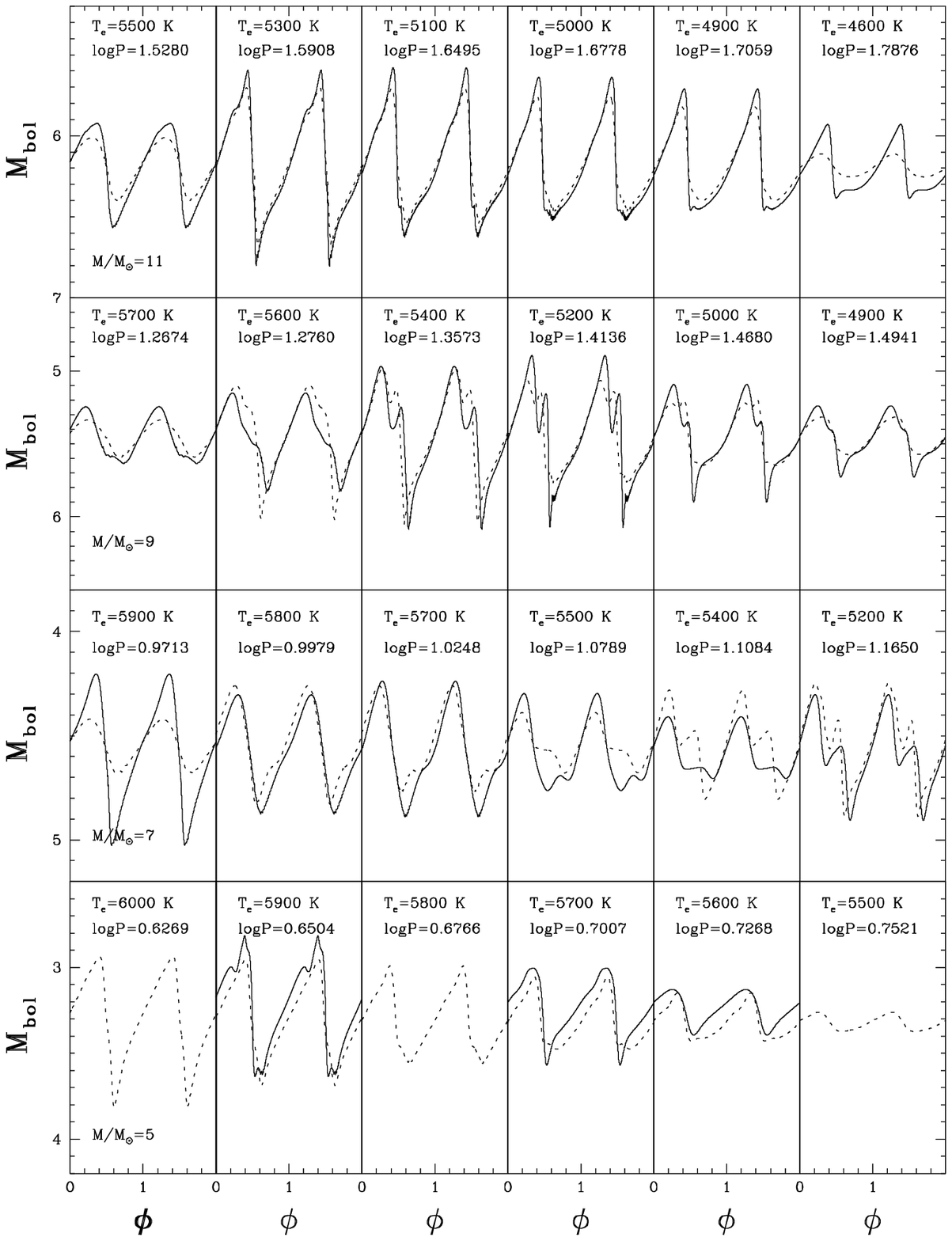}
\caption{\small{Effects of varying $l/H_p$ from 1.5 (solid lines) to
1.8 (dashed lines) on fundamental bolometric light curves at
$Z$=0.004.}} \label{7587fig3}
\end{figure}

The bolometric light curves of selected models\footnote{All the light
and radial velocity curves are available upon request to
giuliana.fiorentino@oabo.inaf.it.} with $Z$=0.02, $Y$=0.28 and with
$Z$=0.004, $Y$=0.25 are presented respectively in Figs. 2 and 3. As
discussed in Paper I and Paper III, the morphology and the amplitude
of the curves vary within the instability strip, depending on the
adopted chemical composition, mass and luminosity. In particular,
variations of the stellar mass for a fixed $ML$ relation or variations
in the $ML$ relation for a fixed mass, affect both the amplitude and
the morphology of the light curves (see discussion in Paper I and
Paper III). As for the dependence on the mixing length parameter, we
notice that increasing the $l/H_p$ value the pulsation amplitude and
the size of the various secondary features decrease.  Concerning the
Hertzsprung progression (HP), which is the relationship between the
pulsation period and the phase of the bump that appears in the light
and radial velocity curves for periods around 10 d, we find that in
the range $0.004\le Z\le 0.02$ an increase in the metal content causes
a shift of the HP center toward shorter periods (see also Paper VI and
Paper IX) for each selected $l/H_p$ value.

\subsection{Predicted Period-Wesenheit relations}

The bolometric light curves provided
by the non-linear approach have been transformed into the
observational bands $UBVRIJK$ by means of the model atmospheres by
Castelli, Gratton \& Kurucz (1997a,b). Then, the amplitudes in the
various spectral bands and, after a time integration,  the pulsator predicted
magnitude- and intensity-averaged mean magnitudes are derived. In
the following, we will refer to selected intensity-weighted mean magnitudes
$\langle M_i\rangle$ and colors [$\langle M_i\rangle -\langle
M_j\rangle$], but the whole set of data, including
magnitude-averaged values, are available upon request to the
authors.

%fig4
\begin{figure}
\includegraphics[width=8cm]{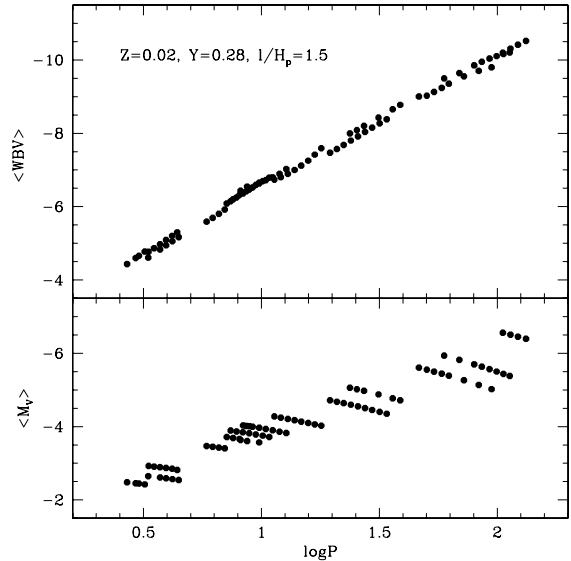}
\caption{\small{Absolute visual magnitude $\langle M_V\rangle$
and $WBV$ function versus period
for fundamental pulsators with $Z$=0.02, $Y$=0.28 and $l/H_p$=1.5.}}
\label{7587fig4}
\end{figure}

%fig5
\begin{figure}
\includegraphics[width=8cm]{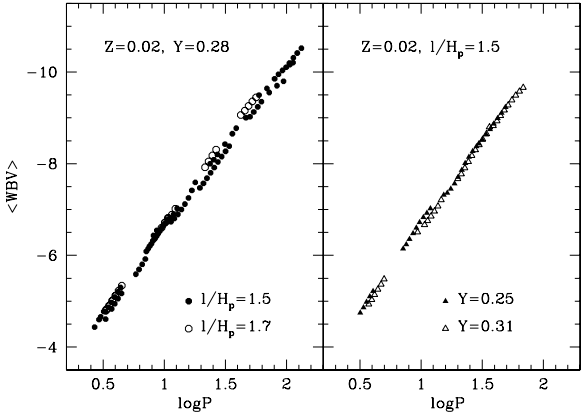}
\caption{\small{$WBV$ function versus period for fundamental pulsators with
$Z$=0.02, varying the value of the mixing-length parameter (left
panel) and the helium content (right panel).}}
\label{7587fig5}
\end{figure}

%fig6
\begin{figure}
\includegraphics[width=8cm]{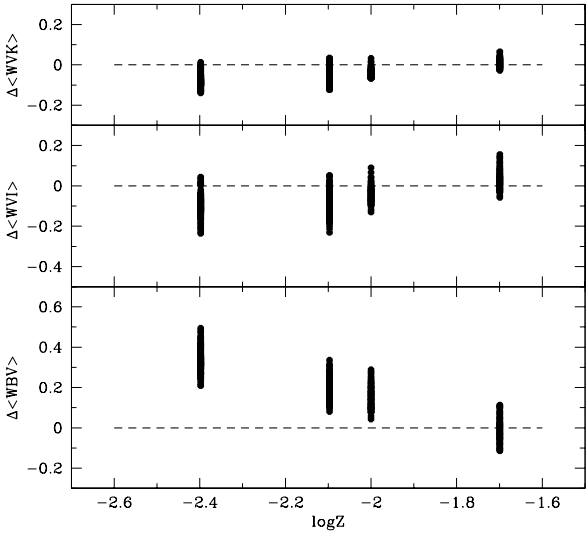}
\caption{\small{Residuals of the Wesenheit functions with respect to
the $PW$ relations at $Z$=0.02, $Y$=0.28 and $l/H_p$=1.5.}}
\label{7587fig6}
\end{figure}

It is well known that a restatement of the Stefan's law for
pulsating stars is the period relation $P$=$f(M,L,T_e)$ for
which the pulsation period of each given variable is uniquely
defined by its mass, bolometric luminosity, and effective
temperature. Once transformed into the observational plane,
this relation yields that the pulsator absolute magnitude $M_i$
in a given photometric bandpass is a linear function of the
pulsation period, mass ($M$ in solar units) and color index
[$\langle M_i\rangle -\langle M_j\rangle$], i.e.,
$$M_i = a+b\log P+c\log M+d[\langle M_i\rangle -\langle M_j\rangle],\eqno(1)$$
\noindent for each given chemical composition. We wish to emphasize
that the mass-dependent Period-Luminosity-Color ($PLC$) relations
hold for individual variables and their use together with measured
metal contents, absolute magnitudes and intrinsic colors provides
precise values of the ``pulsation'' mass $M_p$, i.e. the actual
mass, of each Cepheid.  In the case of Cepheids located at the same
distance and with the same reddening and metal content, one can
estimate the mass range covered by the variables, and in turn the
slope of the empirical $M_pL$ relation, independently of the
distance and reddening value {\bf (see also Beaulieu, Buchler \& Kolláth 2001)}.

However, the adoption of the $PLC$ relations requires accurate
knowledge of the reddening and for this reason, {\bf taking
advantage of the evidence that the color coefficients are not too
different from the extinction-to-reddening ratios provided by
optical and near-infrared reddening laws, the so-named Wesenheit
functions have been widely used to bypass interstellar extinction
problems (see Madore 1982; Madore \& Freedman 1991; Tanvir 1999;
Paper V)}.  In this paper, adopting from Dean, Warren \& Cousins
(1978) and Laney \& Stobie (1993) the absorption to reddening ratios
$A_\lambda/E(B-V)$
$$A_V=3.07+0.28(B-V)_0+0.04E(B-V)$$
$$A_I=1.82+0.205(B-V)_0+0.022E(B-V)$$
$$A_J=0.764; A_K=0.279$$
\noindent we use\footnote{By
artificially reddening our pulsation models, we
find $A_V/E(B-V)$=3.30$\pm$0.05 and
$A_I/E(B-V)$=1.99$\pm$0.04 with $E(B-V)$ up to 2 mag.}
the quantities
$WBV=\langle M_V\rangle -$3.30[$\langle M_B\rangle
-\langle M_V\rangle$], $WVI=\langle M_V\rangle -$2.52[$\langle
M_V\rangle -\langle M_I\rangle$], $WVJ=\langle M_V\rangle -$1.30[$\langle M_V\rangle
-\langle M_J\rangle$], $WVK=\langle M_V\rangle -$1.09[$\langle
M_V\rangle -\langle M_K\rangle$] and $WJK=\langle M_K\rangle -$0.575[$\langle
M_J\rangle -\langle M_K\rangle$].

As shown in Fig. 4 for a selected chemical composition, the
adoption of the reddening-free Wesenheit functions has also the
effect to significantly reduce the dispersion of visual magnitudes
at a given period. Moreover, we wish to stress that the $PW$
relations have the undeniable advantage to be {\it linear} over
the whole period range, as already discussed in Paper V and
recently emphasized by Ngeow \& Kanbur (2005) from measurements of
LMC Cepheids. Regarding the effects of $\l/H_p$ (at fixed $Z$ and
$Y$) and $Y$ (at fixed $\l/H_p$ and $Z$), we show in Fig. 5 that
they are quite negligible.  Conversely, adopting the $PW$
relations at $Z$=0.02, $Y$=0.28 and $l/H_p$=1.5 as reference
relations, we show in Fig. 6 that the residuals of all the models
listed in Table 1 suggest a metallicity effect in the sense that
the $WBV$ quantity becomes fainter whereas $WVI$ and $WVK$ become
slightly brighter as the metal content decreases, at constant
period.

\begin{table*}\label{wes}
\caption{Predicted mass-dependent $PW$ relations for fundamental
pulsators with $Z$=0.004-0.02, as based on intensity-averaged
magnitudes of the pulsators. The intrinsic dispersion of the
constant term includes the variation of helium content and
mixing-length parameter, at constant $Z$. The last column gives the
resulting uncertainty $\epsilon_{PW}$(log$M_p$) on the pulsation
mass inferred by these relations.}
\begin{center}
\begin{tabular}{lccccc}
\hline \hline
$W$ & $a$ & $b$  & $c$ & $d$ & $\epsilon_{PW}$(log$M_p$)\\
\hline \multicolumn{6}{c}{$W$=$a$+$b$log$P$+$c$log$M_p$+ $d$log$Z$}
\\
$WBV$&$-3.52\pm$0.07&$-3.45\pm$0.02&$-1.16\pm$0.05&$-0.67\pm$0.01& 0.06\\
$WVI$&$-1.21\pm$0.05&$-2.43\pm$0.01&$-2.66\pm$0.04&$+0.14\pm$0.01& 0.02\\
$WVJ$&$-0.79\pm$0.05&$-2.30\pm$0.01&$-2.65\pm$0.04&$+0.17\pm$0.01& 0.02\\
$WVK$&$-1.08\pm$0.04&$-2.47\pm$0.01& $-2.64\pm$0.03&$+0.15\pm$0.01& 0.02\\
$WJK$&$-1.20\pm$0.04&$-2.55\pm$0.01& $-2.55\pm$0.03&$+0.14\pm$0.01& 0.02\\
\hline
\end{tabular}
\end{center}
%\end{flushleft}
\end{table*}

\begin{table*}\label{wese}
\caption{Predicted evolutionary $PW_e$ relations for fundamental
pulsators with $Z$=0.004-0.02, as based on intensity-averaged
magnitudes of the pulsators. The last column gives the intrinsic
dispersion of the relations. }

\begin{center}
\begin{tabular}{lccccc}
\hline \hline
$W$ & $\alpha$ & $\beta$  & $\gamma$ & $\delta$ & $\sigma_W$ \\
\hline
\multicolumn{6}{c}{$W$=$\alpha$+$\beta$log$P$+$\gamma$log$L/L_{can}$+$\delta$log$Z$}
\\
$WBV$&$-4.18\pm$0.06&$-3.84\pm$0.01&$+0.72\pm$0.02&$-0.71\pm$0.01 & 0.06\\
$WVI$&$-2.67\pm$0.11& $-3.30\pm$0.01&$+0.84\pm$0.04&$+0.08\pm$0.01 & 0.11 \\
$WVJ$&$-2.36\pm$0.12& $-3.23\pm$0.01&$+0.86\pm$0.04&$+0.11\pm$0.01 & 0.12\\
$WVK$&$-2.53\pm$0.10& $-3.33\pm$0.01&$+0.89\pm$0.04&$+0.09\pm$0.01 & 0.10\\
$WJK$&$-2.61\pm$0.09& $-3.38\pm$0.01&$+0.90\pm$0.03&$+0.08\pm$0.01 & 0.10\\
\hline
\end{tabular}
\end{center}
\end{table*}

As a whole, taking the pulsator mass and luminosity as independent
parameters, a linear interpolation through all the fundamental
models listed in Table 1 yields the mass-dependent $PW$ relations
given in Table 2, where the intrinsic dispersion of the constant
term includes the variation of $Y$ and $l/H_p$ at constant $Z$.
These $PW$ relations cannot be used to get distances but are
important to estimate, independently of reddening, the Cepheid
pulsation mass $M_p$ with the formal uncertainty
$\epsilon_{PW}$(log$M_p$) listed in the last column in this Table,
once the distance and metal content are known.

We said that in the framework of canonical (no mass-loss, no
convective core overshooting) evolutionary predictions, the
bolometric luminosity of a Classical Cepheid is mainly governed by
the chemical composition and the original, or ``evolutionary'',
stellar mass $M_e$. The properties of central He-burning
intermediate-mass stars have been extensively discussed (see CCS;
B00; Girardi et al. 2000; Castellani et al. 2003 and references
therein) and the relevant literature contains several theoretical
$ML$ relations which are widely used to estimate the Cepheid
evolutionary mass.

In this paper, as reference canonical $ML$
relation in the mass range 4-15$M_{\odot}$, we adopt the B00 relation:
$$\log L_{can}=0.90+3.35\log M_e+1.36\log Y-0.34\log Z\eqno(2)$$

\noindent where mass and luminosity are in solar units. This
relation has a standard deviation $\sigma$=0.02 which accounts for
both the blueward and the redward portion of the loop, i.e., the
2$^{nd}$ and 3$^{rd}$ crossing\footnote{The first crossing of the
instability strip deals with the post-ZAMS evolution towards the
Red Giant Branch.}, respectively, of the pulsation region. The
introduction of a $ML$ relation, which bounds the allowed values
of mass and luminosity, yields ``evolutionary'' $PW_e$ relations
where the mass-term is removed. This procedure has been followed
in our previous papers, whereas here we wish to avoid any
assumption on the $ML$ relation taking advantage of the various
luminosity levels listed in Table 1 for a fixed mass.

Based on the canonical luminosity given by Eq. (2) as a reference
level, we estimate the difference log$L/L_{can}$ for each
pulsation model. Then, by a linear interpolation through all the
models we get the relations given in Table 3. According to these
relations, once the value of the $L/L_{can}$ ratio is adopted,
where $L_{can}$ is given by Eq. (2), we can determine the distance
to individual Cepheids with the accuracy $\sigma_W$ listed in the
last column of the Table, provided that the Cepheid metal content
is known. Let us emphasize that, aside from minor discrepancies
among the canonical $ML$ relations provided by different authors,
a variation in luminosity with respect to the reference canonical
value can originate in the occurrence of a convective core
overshooting or mass-loss before or during the He-burning phase.
In the former case, the star is over-luminous by
log$L/L_{can}\sim$0.20 for its mass and chemical composition (see
also Chiosi, Wood \& Capitanio 1993), in the latter, the star has
a smaller mass for its luminosity and, depending on the progenitor
mass $M_{pr}$ and the amount of mass-loss $\Delta M$, one can
estimate log$L/L_{can}\sim$0.9$\Delta M/M_{pr}$ (see Castellani \&
Degl'Innocenti 1995; B00; Castellani et al. 2003). Of importance,
is that both cases yield {\it positive} log$L/L_{can}$ ratios and
that the $PW_e$ relations at $L=L_{can}$ provide the {\it maximum}
value of the Cepheid distance, with the intrinsic distance modulus
decreasing as $\delta\mu_0=\gamma$log$L/L_{can}$, where $\gamma$
is the coefficient listed in Table 3. Finally, we wish to draw
attention on the opposite metallicity dependence of the $P$-$WBV$
relation with respect to the other Wesenheit functions. Such a
feature provides a quite straight method to estimate the Cepheid
metal content, independently of distance and reddening and with a
small effect of the adopted $L/L_{can}$ ratio (see Sect. 4).

\section{Synthetic PL relations}

In the previous section, it has been shown that the $PW$ relations
are quite unaffected by the value of the mixing-length parameter,
although the edges of the pulsation region with $Z\ge$ 0.008 depend
on the adopted $l/H_p$ value. In order to investigate the role of
the mixing-length parameter on the $PL$ relations, we populate the
predicted instability strip adopting evolutionary computations and
using the procedure early suggested by Kennicutt et al. (1998) and
adopted in our previous papers.

In Paper V and Paper IX, we have already derived synthetic $PL$
relations by populating the predicted instability strip according to
an assumed mass distribution and adopting, for each mass and
chemical composition, the canonical luminosity given by Eq. (2).
Here, we present new synthetic $PL$ relations that use evolutionary
tracks and take into account also the evolutionary times spent by
the Cepheids inside the strip. To this purpose, we adopt the
evolutionary computations by
 Pietrinferni et al. (2004) that cover a mass
range from 3 to 10$M_{\odot}$ and a very large metallicity range.
Before proceeding, let us note that the mixing-length parameter used
in the evolutionary computations is $l/H_p$=1.9. However, the
meaning of this parameter when computing evolutionary sequences is
different from the one adopted in the pulsation code. The discussion
of the differences between the two convective treatments is beyond
the scope of this paper and for details we refer the interested
readers to Bono \& Stellingwerf (1994) and Stellingwerf (1982).

To build a synthetic population we extract each mass by using an
Initial Mass Function (IMF) with a power law distribution ($P(M)
\sim 1/M^3$) and a constant Star Formation Rate. Even if the
latter assumption may be not reliable for a complex stellar
population like LMC (Cioni et al. 2006), our method of populating
stellar tracks crossing the IS is an improvement of the classical
uniform distribution. By interpolating the evolutionary tracks, we
assign both a luminosity and an effective temperature to each
synthetic star, and the total number of extractions ($\sim$ 50000)
is fixed in order to obtain a number of synthetic stars falling
into the instability strip of the order of $10^3$, in analogy to
our previous procedures. Once built a synthetic stellar
population, we use the constraints of the pulsation theory to
reject objects outside the strip and to assign the expected period
to each selected fundamental pulsator. To this purpose, we {\bf
interpolate through} the predicted boundary temperatures (FBE and
FRE, see columns 7 and 8 in Table 1) in order to select the
synthetic stars with FBE$\ge$log$T_e\ge$ FRE and we determine the
period by means of the $P=f(L,M,T_e,Z)$ relation provided by a
regression through all the fundamental models. {\bf Then, using
the same atmosphere models adopted in Sect. 2.2, we calculate the
pulsator absolute magnitude in the various photometric bands. Left
and right panels in Fig. 7 show the results of these simulations
at $Z$=0.004 and 0.02, respectively, adopting $l/H_p$=1.5. It is
of interest to note that decreasing the metal content increases
the predicted number of faint and hot (i.e., short period)
pulsators, fully supporting recent suggestions (see Cordier,
Goupil, Lebreton 2003) that explain the occurrence of faint short
period Cepheids in the Small Magellanic Cloud ($Z\sim$ 0.004) as
an effect of the decreased metallicity.

%fig7
\hspace{-2cm}
\begin{figure}
\caption{\small{Synthetic fundamental pulsators with
$Z$=0.004 and $Y$=0.251 (left panels) and $Z$=0.02, $Y$=0.289
(right panel) adopting $l/H_p$=1.5. The solid and the dashed lines refer to
the linear and the quadratic fits, respectively.}} \label{7587fig7}
\end{figure}

\begin{table*}\label{pllin}
\caption{Synthetic linear $PL$ relations for fundamental
pulsators with log$P\ge$ 0.4 in the
form $M_i$=$\alpha$+$\beta$log$P$. }
\begin{center}
\vspace{0.5truecm}
\begin{tabular}{lcccccccc}
\hline \hline
$M_i$ & $\l/H_p$ & $\alpha$ & $\beta$ &$\sigma$ & $\l/H_p$ & $\alpha$ & $\beta$ &$\sigma$\\
\hline
\multicolumn{9}{c}{$Z$=0.004,$Y$=0.251}\\
$M_B$&1.5& $-0.78\pm$0.02 & $-2.64\pm$0.03 & 0.21 & 1.8 & -- & --&--\\
$M_V$&1.5& $-1.22\pm$0.02 & $-2.84\pm$0.02 & 0.16 & 1.8 & -- & --&--\\
$M_R$&1.5& $-1.52\pm$0.01 & $-2.92\pm$0.02 & 0.14 & 1.8 & -- & --&--\\
$M_I$&1.5& $-1.83\pm$0.01 & $-2.99\pm$0.02 & 0.13 & 1.8 & -- & --&--\\
$M_J$&1.5& $-2.20\pm$0.01 & $-3.09\pm$0.02 & 0.11 & 1.8 & -- & --&--\\
$M_K$&1.5& $-2.55\pm$0.01 & $-3.19\pm$0.01 & 0.10 & 1.8 & -- & --&--\\
\hline
\multicolumn{9}{c}{$Z$=0.008,$Y$=0.251}\\
$M_B$&1.5& $-0.94\pm$0.02 & $-2.37\pm$0.02 & 0.29 & 1.8 & $-0.77\pm$0.02 & $-2.65\pm$0.02 & 0.23\\
$M_V$&1.5& $-1.35\pm$0.02 & $-2.73\pm$0.02 & 0.21 & 1.8 & $-1.22\pm$0.02 & $-2.94\pm$0.02 & 0.17\\
$M_R$&1.5& $-1.62\pm$0.01 & $-2.87\pm$0.02 & 0.18 & 1.8 & $-1.52\pm$0.01 & $-3.04\pm$0.01 & 0.14\\
$M_I$&1.5& $-1.91\pm$0.01 & $-3.00\pm$0.01 & 0.15 & 1.8 & $-1.82\pm$0.01 & $-3.15\pm$0.01 & 0.12\\
$M_J$&1.5& $-2.24\pm$0.01 & $-3.18\pm$0.01 & 0.11 & 1.8 & $-2.18\pm$0.01 & $-3.29\pm$0.01 & 0.09\\
$M_K$&1.5& $-2.54\pm$0.01 & $-3.35\pm$0.01 & 0.07 & 1.8 & $-2.51\pm$0.01 & $-3.41\pm$0.01 & 0.06\\
\hline
\multicolumn{9}{c}{$Z$=0.02,$Y$=0.289}\\
$M_B$&1.5& $-0.90\pm$0.01 & $-2.05\pm$0.02 & 0.16&1.7 & $-0.90\pm$0.02 & $-2.14\pm$0.03 &0.18\\
$M_V$&1.5& $-1.41\pm$0.01 & $-2.48\pm$0.02 & 0.11&1.7 & $-1.41\pm$0.01 & $-2.53\pm$0.02 &0.13\\
$M_R$&1.5& $-1.70\pm$0.01 & $-2.65\pm$0.01 & 0.10&1.7 & $-1.70\pm$0.01 & $-2.69\pm$0.02 &0.11\\
$M_I$&1.5& $-2.00\pm$0.01 & $-2.78\pm$0.01 & 0.08&1.7 & $-2.00\pm$0.01 & $-2.82\pm$0.02 &0.09\\
$M_J$&1.5& $-2.32\pm$0.01 & $-3.03\pm$0.01 & 0.06&1.7 & $-2.33\pm$0.01 & $-3.04\pm$0.01 &0.06\\
$M_K$&1.5& $-2.62\pm$0.01 & $-3.22\pm$0.01 & 0.04&1.7 & $-2.64\pm$0.01 & $-3.21\pm$0.01 &0.04\\
\hline
\end{tabular}
\end{center}
%\end{flushleft}
\end{table*}

\begin{table*}\label{plnonlin}
\caption{Synthetic quadratic $PL$ relations for fundamental
pulsators with log$P\ge$ 0.4 in the
form $M_i$=$\alpha$+$\beta$log$P$+$\gamma$log$P^2$. {\bf Note that the for each relation
the rms dispersion is included in the error of the $\alpha$ coefficient.}}
\begin{center}
\vspace{0.5truecm}
\begin{tabular}{lcccccccc}
\hline \hline
$M_i$ & $\l/H_p$ & $\alpha$ & $\beta$ & $\gamma$ &$\l/H_p$ & $\alpha$ & $\beta$ & $\gamma$\\
\hline
\multicolumn{9}{c}{$Z$=0.004,$Y$=0.251}\\
$M_B$&1.5& $-0.41\pm$0.10 & $-3.64\pm$0.15 & $+0.60\pm$0.09 & 1.8 & -- & --\\
$M_V$&1.5& $-0.95\pm$0.15 & $-3.59\pm$0.12 & $+0.46\pm$0.07 & 1.8 & -- & --\\
$M_R$&1.5& $-1.27\pm$0.13 & $-3.58\pm$0.11 & $+0.40\pm$0.06 & 1.8 & -- & --\\
$M_I$&1.5& $-1.62\pm$0.11 & $-3.58\pm$0.10 & $+0.35\pm$0.06 & 1.8 & -- & --\\
$M_J$&1.5& $-2.03\pm$0.08 & $-3.57\pm$0.08 & $+0.29\pm$0.05 & 1.8 & -- & --\\
$M_K$&1.5& $-2.41\pm$0.06 & $-3.57\pm$0.08 & $+0.23\pm$0.05 & 1.8 & -- & --\\
\hline
\multicolumn{9}{c}{$Z$=0.008,$Y$=0.251}\\
$M_B$&1.5& $-0.21\pm$0.13 & $-4.26\pm$0.15 & $+1.04\pm$0.08 & 1.8 &  $+0.09\pm$0.10 & $-4.92\pm$0.14 & $+1.28\pm$0.07 \\
$M_V$&1.5& $-0.83\pm$0.20 & $-4.08\pm$0.11 & $+0.74\pm$0.06 & 1.8 &  $-0.61\pm$0.16 & $-4.56\pm$0.10 & $+0.91\pm$0.06 \\
$M_R$&1.5& $-1.18\pm$0.17 & $-4.00\pm$0.09 & $+0.63\pm$0.05 & 1.8 &  $-1.00\pm$0.13 & $-4.41\pm$0.08 & $+0.77\pm$0.05 \\
$M_I$&1.5& $-1.54\pm$0.14 & $-3.95\pm$0.08 & $+0.52\pm$0.04 & 1.8 &  $-1.39\pm$0.11 & $-4.29\pm$0.07 & $+0.64\pm$0.04 \\
$M_J$&1.5& $-1.98\pm$0.10 & $-3.85\pm$0.06 & $+0.37\pm$0.03 & 1.8 &  $-1.87\pm$0.08 & $-4.10\pm$0.05 & $+0.46\pm$0.03 \\
$M_K$&1.5& $-2.38\pm$0.06 & $-3.77\pm$0.04 & $+0.23\pm$0.02 & 1.8 &  $-2.31\pm$0.05 & $-3.93\pm$0.04 & $+0.29\pm$0.02 \\
\hline
\multicolumn{9}{c}{$Z$=0.02,$Y$=0.289}\\
$M_B$&1.5& $-0.55\pm$0.08 & $-3.09\pm$0.14 & $+0.66\pm$0.09 & 1.7 & $-0.69\pm$0.10 & $-2.79\pm$0.18 & $+0.45\pm$0.12 \\
$M_V$&1.5& $-1.19\pm$0.12 & $-3.12\pm$0.10 & $+0.41\pm$0.06 & 1.7 & $-1.30\pm$0.16 & $-2.87\pm$0.13 & $+0.23\pm$0.09 \\
$M_R$&1.5& $-1.54\pm$0.10 & $-3.13\pm$0.09 & $+0.31\pm$0.05 & 1.7 & $-1.64\pm$0.13 & $-2.90\pm$0.11 & $+0.14\pm$0.08 \\
$M_I$&1.5& $-1.87\pm$0.09 & $-3.15\pm$0.07 & $+0.23\pm$0.05 & 1.7 & $-1.96\pm$0.11 & $-2.94\pm$0.09 & $+0.08\pm$0.06 \\
$M_J$&1.5& $-2.29\pm$0.06 & $-3.10\pm$0.05 & $+0.04\pm$0.03 & 1.7 & $-2.36\pm$0.07 & $-2.94\pm$0.06 & $-0.07\pm$0.04 \\
$M_K$&1.5& $-2.67\pm$0.04 & $-3.09\pm$0.04 & $-0.09\pm$0.02 & 1.7 & $-2.71\pm$0.04 & $-2.98\pm$0.04 & $-0.16\pm$0.03 \\
\hline
\end{tabular}
\end{center}
%\end{flushleft}
\end{table*}

For the sake of the following comparison with observed Cepheids,
we select the predicted fundamental pulsators with log$P\ge$ 0.4
and, by a regression through the synthetic populations we get the
$PL$ relations listed in Table 4 (linear approximation) and Table
5 (quadratic approximation).} One can see that the $l/H_p$
variation from 1.5 to 1.7 produces minor effects on both the slope
and zero point, at fixed chemical composition. More importantly,
present $PL$ relations confirm most of our previous results. In
particular, we find that:
\begin{enumerate}
\item the pulsator distributions are better described
by quadratic relations, mainly in the optical bands;
\item {\bf the slope and the intrinsic dispersion of the predicted Period-Magnitude
distribution at fixed metal content decrease moving from optical to
near-infrared bands, in
agreement with well-known empirical results (see, e.g., Madore \&
Freedman 1991)}.
\item decreasing the metal content, the linear $PL$ relations become steeper,
depending on the filter wavelength: from $Z$=0.02 to
0.004 the slope of the $P$-$M_B$, $P$-$M_V$ and $P$-$M_I$ relations
vary by $\sim$29\%, 15\% and $\sim$8\%, respectively, with no
significant effects on the near-infrared relations;
\item regarding the zero-point of the $PL$ relations,
{\it at periods longer than $\sim$ 4-5 days} the metal-poor
pulsators have brighter optical magnitudes than the metal-rich ones, again depending on
the photometric band. Varying the metal content from $Z$=0.02 to
0.004, the $B$, $V$ and $I$ magnitudes at log$P$=1 become brighter by
$\sim$0.47, $\sim$0.18 and $\sim$0.06 mag, whereas the $K$ magnitudes
become fainter by $\sim$0.10 mag.
\end{enumerate}

Finally, by comparing the new results at $l/H_p$=1.5 with those
presented in Paper V, we can estimate the effects of different
methods to derive the synthetic $PL$ relations. As a whole, the new linear
slopes appear somehow steeper ($\sim$ 6\%) than our previous ones,
while the predicted magnitudes at log$P$=1 agree within $\pm$ 0.1
mag.

\section{Theory versus observations}

In order to test the predictive potential of our pulsation scenario,
we compare in the following the theoretical relations discussed in
the previous sections with selected samples of Galactic and Magellanic Cloud
Cepheids.

\subsection{Galactic Cepheids with absolute trigonometric parallaxes}

The absolute $VIK$ magnitudes determined by Benedict et al. (2007,
hereafter B07) on the basis of $HST$ trigonometric parallaxes
yield the $PL$ relations ($M_i$=$\alpha$+$\beta$log$P$) reported
in Table 6. {\bf Based on these absolute magnitudes and adopting
$A_V/E(B-V)$=3.3 and $A_V/E(V-I)$=2.52, we derive the $PW$
relations listed in the same Table. Note that the $P$-$WBV$
relation is determined by using the $(B-V)$ and $E(B-V)$
values given by B07.}

%table 6
\begin{table}\label{tab6}
\caption{Absolute $PL$ and $PW$ relations for fundamental Galactic Cepheids with
measured distance. The near-infrared magnitudes are in the CIT
photometric system.}
\begin{center}
\vspace{0.5truecm}
\begin{tabular}{lcclcc}
\hline \hline
$M_i$ & $\alpha$ & $\beta$ & $W$ & $\alpha$ & $\beta$  \\
\hline \multicolumn{6}{c}{B07}
\\
$M_V$ & $-$1.62 & $-$2.43 & $WBV$ & $-$2.03 & $-$4.52\\
$M_I$ & $-$1.97 & $-$2.81 & $WVI$ & $-$2.47 & $-$3.43\\
$M_K$ & $-$2.39 & $-$3.32 & $WVK$ & $-$2.45 & $-$3.43 \\\\
 \multicolumn{6}{c}{S04}
\\
$M_V$ & $-$0.95 & $-$3.08 & $WBV$ & $-$2.04 & $-$4.35\\
$M_I$ & $-$1.49 & $-$3.30 & $WVI$ & $-$2.11 & $-$3.78\\
$M_J$ & $-$1.77 & $-$3.53 & $WVK$ & $-$2.00 & $-$3.78\\
$M_K$ & $-$2.02 & $-$3.67 & $WJK$ & $-$2.08 & $-$3.80\\
\hline
\end{tabular}
\end{center}
\end{table}

Bearing in mind that the $PL$ relations, {\bf mostly in the
optical bands, depend on the finite width of the pulsation region
and on} the stellar distribution within the instability strip, the
observed slope and zero-point provided by the B07 Cepheids appear
comfortably consistent with the predicted values at $Z$=0.02
presented in Table 4. It is also worth noticing that, compared
with the results given in Table 11 for LMC ($Z$=0.008) Cepheids,
as determined by hundreds of variables observed by the OGLE team
and Persson and coworkers (see later), the Galactic $PL$ relations
in the $VI$ bands have a milder slope, in agreement with the
predicted behavior. Regarding the $PW$ relations, which are less
affected by the finite width of the instability strip, the
Galactic $P$-$WVI$ and $P$-$WVK$ slopes differ by only $\sim$ 5\%
from the predicted values at constant metal content and
$L/L_{can}$ ratio (see Table 3), whereas the observed $P$-$WBV$
relation is statistically steeper than the predicted one.

In order to investigate this discrepancy, in the following we
analyze the Cepheid apparent magnitudes of the B07 sample to show
the capabilities of the predicted relations to get useful
information on several stellar parameters, as summarized in Table
7. {\bf Note that the last line in this Table deals with the case
that FF Aql is an overtone pulsator (see Antonello, Poretti \&
Reduzzi 1990; Kienzle et al. 1999). Accordingly, the observed
period has been increased by $\delta$log$P$=0.156, as suggested by
the pulsation theory (see also Alcock et al. 1995), in order to
use our predicted fundamental relations.}

As a first step, we adopt log$L/L_{can}$=0 (canonical evolutionary
frame) and the solar metallicity $Z_{\odot}$=0.017 to estimate the
``pulsation'' distance moduli $\mu_0$ from the predicted $PW_e$
relations and the observed $WBV$, $WVI$ and $WVK$ quantities. Note
that our models adopt the Bessell \& Brett (1988) near-infrared
system and for this reason the $K$(CIT) magnitudes have been
increased by 0.02 mag according to Carpenter (2001). As shown in
the columns (4)-(6) in Table 7, the distances given by the $WBV$
functions are generally longer than the $WVI$ and $WVK$-based
values, with the only exception of l Car which shows an opposite
behaviour. Then, by {\bf exploiting the metallicity dependence of
the various $PW_e$ relations (see Table 3), we demand that $WBV$,
$WVI$ and $WVK$ provide the same distance and} we eventually
derive the $Z$ and $\langle\mu_{0,W}\rangle$ values listed in
column (7) and (8), respectively, while the following two columns
give the results inferred by repeating the procedure with
log$L/L_{can}$=0.2.

%fig8
\begin{figure}
\includegraphics[width=8cm]{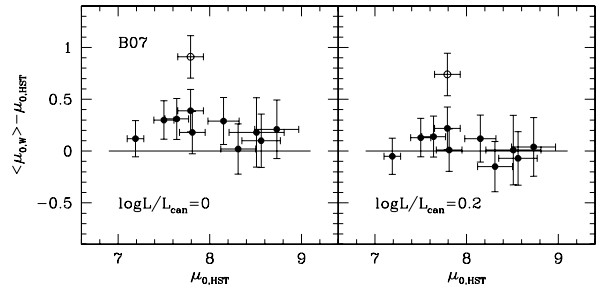}
\caption{\small{Pulsation distance moduli of
Galactic Cepheids versus $HST$-based values under two assumptions of
the log$L/L_{can}$ ratio. The open circle refers to FF Aql, if it
were an overtone pulsator (see text).}} \label{7587fig8}
\end{figure}

{\bf As a whole, we get metal abundances quite consistent with a solar-like value and
pulsation distances in statistical agreement
with the parallax-based values, closer for the non-canonical value log$L/L_{can}$=0.2
(see Fig. 8). However, since
$\delta \langle\mu_{0,W}\rangle$=$-0.85\delta$log$L/L_{can}$,
we take the $HST$-based distance in column (3) at face value to
derive the final log$L/L_{can}$ and $Z$ values of each
individual variable, as listed in the last two columns in the Table.
For all the variables we find positive
log$L/L_{can}$ ratios which suggest an average value of about 0.25
or a fair period dependence as log$L/L_{can}$=0.41$-$0.19log$P$.
Regarding FF Aql, if it were an overtone pulsator, it would have an exceedingly
large log$L/L_{can}$ ratio to fit the
$HST$-based distance, leading us to conclude that this variable
is a fundamental pulsator.}

All the variables yield log$Z=-$2.0$\pm$0.1, with the only
exception of l Car which seems to be significantly more
metal-rich. {\bf Indeed, the high-resolution spectroscopic
measurements by Romaniello et al. (2005) and Lemasle et al. (2007)
for $\beta$ Dor ([Fe/H]=$-$0.14), $\zeta$ Gem ([Fe/H]=$-$0.19) and
l Car ([Fe/H]=+0.10) seem to support the metal overabundance of l
Car with respect to the other variables. It is of interest to note
that the exclusion of l Car would actually lead to mildly {\it
steeper} $P$-$M_V$ and $P$-$M_I$ relations, i.e.
$M_V=-1.43-2.67$log$P$ and $M_I=-1.82-3.01$log$P$, but still
shallower than the LMC counterparts (see Table 11). Furthermore,
we note that, with respect to these new relations at [Fe/H]$\sim
-$2.0, l Car ({\bf[Fe/H]=$+$0.10}) turns out to be {\it fainter }
by $\delta M_V$=0.22 mag and $\delta M_I$=0.18 mag, in agreement
with the predicted metallicity effect.

Eventually, we wish to mention that using the $HST$-based distance
in column (3) and the metal content in column (12) together with
the predicted mass-dependent $PW$ relations given in Table 2 we
estimate that the mass of W Sgr is 6.0$\pm$1.5$M_{\odot}$, in
close agreement with the value 6.5$\pm$2$M_{\odot}$ listed by B07.
Regarding FF Aql, we find 3.5$\pm$1.5$M_{\odot}$ (fundamental
pulsator) and 2.3$\pm$1.5$M_{\odot}$ (first overtone) which
compared with the B07 mass 4.5$\pm$1$M_{\odot}$ should again
support that this variable pulsate in the fundamental mode. The
comparison between the evolutionary and the pulsation mass of
Galactic and LMC Cepheids will be discussed in a forthcoming
paper.}

\subsection{Galactic Cepheids with ISB distances}
The absolute magnitudes $BVIJK$ determined by S04 using the
Infrared Surface Brightness (ISB) technique provide the absolute
$PL$ and $PW$ relations reported in Table 6.  These relations are
significantly steeper in comparison with the predicted linear
relations at solar metal content, as well as with respect to the
B07 and the LMC results listed in Table 6 and Table 11,
respectively.

%fig9
\begin{figure}
\includegraphics[width=8cm]{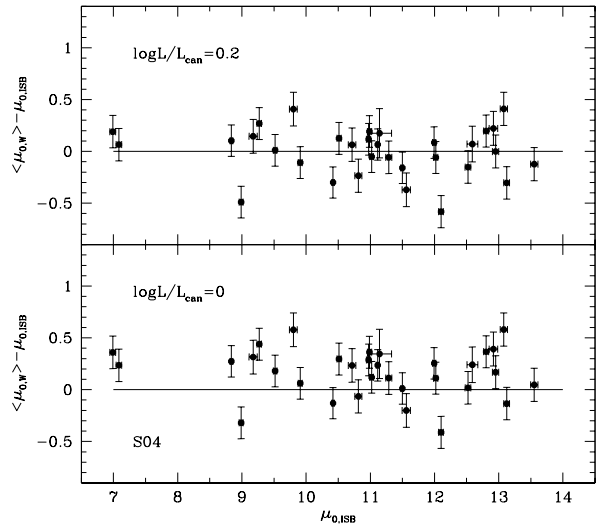}
\caption{\small{Pulsation distance moduli of
Galactic Cepheids versus ISB-based values under two assumptions of
the log$L/L_{can}$ ratio. }} \label{7587fig9}
\end{figure}

By repeating the procedure adopted for the B07 sample, the S04
variables yield the results listed in Table 8, {\bf while Fig. 9
shows that no statistical agreement between pulsation and
ISB-based distances is found with log$L/L_{can}$=0 nor 0.2. As a
fact, based on the $\langle\mu_{0,W}\rangle$-log$L/L_{can}$
relation, we use the ISB-based distances to determine the ratios
listed in column (11) in Table 8 and we draw the attention on the
resulting quite large spread as well as on some unrealistic
negative values towards the longer periods. Moreover, a straight
regression through the results yield
log$L/L_{can}$=0.85$-$0.55log$P$, which is significantly different
from the B07 Cepheid result.}

%fig10
\begin{figure}
\includegraphics[width=8cm]{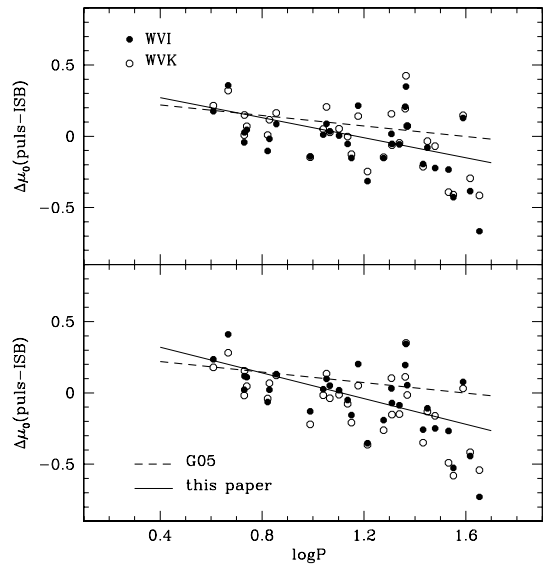}
\caption{\small{(upper panel) - Corrections to ISB-based distances to
S04 Cepheids by using the $P$-$WVI$ and $P$-$WVK$ relations provided
by the B07 variables; (lower panel) - As in the upper panel, but
adopting log$L/L_{can}$=0.25, which is the average value of the B07
Cepheids. The solid line is the linear regression through the points,
while the dashed line is the correction suggested by G05.}}
\label{7587fig10}
\end{figure}

Truly, we should mention that G05 have already shown that the ISB
distances to 13 LMC Cepheids, even corrected for the tilt of the
LMC bar, show a puzzling dependence on the period, with the
shortest period variables giving $\mu_0$(LMC)$\sim$ 18.3 mag and
the longest period ones $\sim$ 18.6 mag ({\bf see the values
listed in column (3) in Table 12), and yield $PL$ relations
significantly steeper (see column (3) in Table 9) than those
determined by the OGLE and P04's huge samples}.  On this ground,
having identified that the adopted p-factor which is used to
convert radial into pulsation velocities is most likely
responsible of this behavior, they suggest a correction to the
former ISB distances as given by
$$\Delta\mu_0(ISB_{new}-ISB_{old})=0.29-0.18\log P\eqno(3)$$
\noindent with a LMC distance modulus $\mu_0$(LMC)=18.56 mag.
Using this correction to re-determine the distances to the LMC and
S04 Galactic Cepheids, they eventually attain the agreement with
the OGLE and P04 $PL$ relations and no significant difference
between the slopes of the various relations between Galactic
and LMC variables (see last two columns in Table 9). {\bf However,
we note that the slope variation ($\Delta\beta$=+0.18) produced by the G05
correction leaves that also the revised $PL$ and $PW$ relations of the
S04 Cepheids are steeper in comparison with the B07 results.
Furthermore, using Eq. (3) to correct the ISB distance moduli of
the S04 Cepheids we find log$L/L_{can}$=0.50$-$0.32log$P$, which
is still different from the result inferred by the B07 Cepheids.

Actually, the comparison between the $HST$ and ISB distances for
$\delta$ Cep (log$P$=0.730) and l Car (log$P$=1.551), the only two
common variables, gives $\Delta\mu_0(HST-ISB_{old})=0.10\pm0.10$ mag
and $-0.43\pm0.21$ mag, respectively, while the revised ISB distances
give $\Delta\mu_0(HST-ISB_{rev})=-0.06\pm0.10$ mag for the
short-period Cepheid and $-0.44\pm0.21$ mag for the long-period one,
suggesting a more significant period-dependent correction to the ISB
distances. On the other hand, it seems reasonable to assume that the
Galactic Cepheids in the two samples follow similar relations and/or
evolutionary prescriptions. On this ground, we show in Fig. 10 that
assuming for the S04 Cepheids the absolute $P$-$WVI$ and $P$-$WVK$
relations provided by the B07 variables the correction is
$$\Delta\mu_0(puls-ISB_{old})=0.41-0.35\log P,\eqno(4)$$
\noindent while adopting log$L/L_{can}$=0.25, which is the average
value of the B07 Cepheids, one has
$$\Delta\mu_0(puls-ISB_{old})=0.50-0.45\log P\eqno(5)$$
\noindent It goes without saying that these two $HST$-based
corrections would lead the $PL$ and $PW$ relations of the S04
Cepheids to agree with the B07 results.}

%table 7
\linespread{0.7}
\begin{table*}\label{tab7}
\caption{Galactic Cepheids with distance based on the $HST$
trigonometric parallax measured by B07. From left
to right, the columns give: name (1), period (2), absolute distance
modulus (mag) based on the $HST$ trigonometric parallax, with the
average associated error (3), and absolute distance moduli
determined by the predicted evolutionary $P$-$WBV$ (4), $P$-$WVI$ (5) and
$P$-$WVK$ (6) relations adopting $Z$=0.017 and log$L/L_{can}$=0. By
imposing that the three $PW$ relations give the same result, we get
the metal content and pulsation distance modulus at log$L/L_{can}$=0
(columns (7) and (8), respectively) and log$L/L_{can}$=0.2 (columns (9)
and (10), respectively). The comparison with the $HST$-based
distance modulus yields the final log$L/L_{can}$ value and metal content
listed in the last two columns.}
\begin{center}
\vspace{0.5truecm}
\begin{tabular}{lccccccccccc}
\hline \hline
 \tiny{Name    } & \tiny{ log$P$ } & \tiny{  $\mu_{0,HST}$    } & \tiny{   $\mu_{0,WBV}$    } & \tiny{   $\mu_{0,WVI}$    } & \tiny{   $\mu_{0,WVK}$    } & \tiny{   log$Z$  } & \tiny{   $\langle\mu_{0,W}\rangle$ } & \tiny{   log$Z$  } & \tiny{   $\langle\mu_{0,W}\rangle$ } & \tiny{   log$L/L_{can}$  } & \tiny{   log$Z$  }\\
         &  & \tiny{  ($\pm$) } & \tiny{ $\pm$0.06   } & \tiny{ $\pm$0.12   } & \tiny{  $\pm$0.10   } & \tiny{   $\pm$0.15   } & \tiny{   $\pm$0.15   } & \tiny{   $\pm$0.15   } & \tiny{   $\pm$0.15   } & \tiny{   ($\pm$)   } & \tiny{   $\pm$0.15   }\\
 \tiny{1   } & \tiny{   2   } & \tiny{   3   } & \tiny{   4   } & \tiny{   5   } & \tiny{   6   } & \tiny{   7   } & \tiny{   8   } & \tiny{   9   } & \tiny{   10  } & \tiny{   11 } & \tiny{12}\\
\hline
 \tiny{ RTAur   } & \tiny{   0.572 } & \tiny{8.15(0.17)} & \tiny{8.62} & \tiny{8.43} & \tiny{8.40} & \tiny{$-$2.02} & \tiny{8.44} & \tiny{$-$2.06} & \tiny{8.27} & \tiny{   +0.34(0.20)} & \tiny{$-$2.09}\\
 \tiny{ TVul    } & \tiny{   0.647 } & \tiny{8.73(0.24)} & \tiny{9.06} & \tiny{8.93} & \tiny{8.91} & \tiny{$-$1.95} & \tiny{8.94} & \tiny{$-$1.98} & \tiny{8.77} & \tiny{   +0.24(0.28)} & \tiny{$-$1.99}\\
 \tiny{ FFAql   } & \tiny{   0.650 } & \tiny{7.79(0.14)} & \tiny{8.30} & \tiny{8.16} & \tiny{8.17} & \tiny{$-$1.94} & \tiny{8.18} & \tiny{$-$1.98} & \tiny{8.01} & \tiny{   +0.46(0.16)} & \tiny{$-$2.02}\\
 \tiny{ $\delta$Cep } & \tiny{0.730} & \tiny{7.19(0.09)} & \tiny{7.52} & \tiny{7.27} & \tiny{7.30} & \tiny{$-$2.06} & \tiny{7.31} & \tiny{$-$2.09} & \tiny{7.14} & \tiny{   +0.14(0.11)} & \tiny{$-$2.08}\\
 \tiny{ YSgr    } & \tiny{   0.761 } & \tiny{8.51(0.30)} & \tiny{8.77} & \tiny{8.73} & \tiny{8.63} & \tiny{$-$1.88} & \tiny{8.69} & \tiny{$-$1.91} & \tiny{8.52} & \tiny{   +0.21(0.35)} & \tiny{$-$1.92}\\
 \tiny{ XSgr    } & \tiny{   0.846 } & \tiny{7.64(0.13)} & \tiny{8.29} & \tiny{7.90} & \tiny{7.90} & \tiny{$-$2.25} & \tiny{7.95} & \tiny{$-$2.29} & \tiny{7.78} & \tiny{   +0.36(0.15)} & \tiny{$-$2.32}\\
 \tiny{ WSgr    } & \tiny{   0.881 } & \tiny{8.31(0.19)} & \tiny{8.51} & \tiny{8.36} & \tiny{8.27} & \tiny{$-$2.02} & \tiny{8.33} & \tiny{$-$2.05} & \tiny{8.16} & \tiny{   +0.03(0.22)} & \tiny{$-$2.02}\\
 \tiny{ $\beta$Dor  } & \tiny{0.993} & \tiny{7.50(0.11)} & \tiny{7.83} & \tiny{7.80} & \tiny{7.80} & \tiny{$-$1.80} & \tiny{7.80} & \tiny{$-$1.84} & \tiny{7.63} & \tiny{   +0.36(0.13)} & \tiny{$-$1.86}\\
 \tiny{ $\zeta$Gem  } & \tiny{1.007} & \tiny{7.81(0.14)} & \tiny{8.07} & \tiny{7.96} & \tiny{8.00} & \tiny{$-$1.88} & \tiny{7.99} & \tiny{$-$1.92} & \tiny{7.82} & \tiny{   +0.21(0.16)} & \tiny{$-$1.92}\\
 \tiny{ lCar    } & \tiny{   1.551 } & \tiny{8.56(0.21)} & \tiny{8.32} & \tiny{8.70} & \tiny{8.71} & \tiny{$-$1.30} & \tiny{8.66} & \tiny{$-$1.33} & \tiny{8.49} & \tiny{   +0.12(0.26)} & \tiny{$-$1.31}\\
 \tiny{ FFAql(FO) } & \tiny{   0.806 } & \tiny{7.79(0.14)} & \tiny{8.90} & \tiny{8.67} & \tiny{8.69}  & \tiny{$-$2.04} & \tiny{8.70} & \tiny{$-$2.08} & \tiny{8.53} & \tiny{   +1.08(0.16)} & \tiny{$-$2.24}\\
\hline
\end{tabular}
\end{center}
\end{table*}

%table 8
\linespread{0.7}
\begin{table*}\label{tab8}
\caption{As in Table 7, but for the Galactic Cepheids with
distance derived by S04 using the Infrared Surface Brightness
(ISB) technique. From left to right, the columns give: name (1),
period (2), absolute distance modulus (mag) based on the ISB
method (3), absolute distance moduli determined by the predicted
evolutionary $P$-$WBV$ (4), $P$-$WVI$ (5) and $P$-$WVK$ (6)
relations adopting $Z$=0.017 and log$L/L_{can}$=0. By imposing
that the three $PW$ relations give the same result, we get the
metal content and pulsation distance modulus at log$L/L_{can}$=0
(columns (7) and (8), respectively) and log$L/L_{can}$=0.2
(columns (9) and (10), respectively). The comparison with the
ISB-based distance modulus yields the final log$L/L_{can}$ value
and metal content listed in the last two columns.}
\begin{center}
\vspace{0.5truecm}

\begin{tabular}{lccccccccccc}
\hline \hline

 \tiny{Name } & \tiny{ log$P$ } & \tiny{$\mu_{0,ISB}$} & \tiny{$\mu_{0,WBV}$} & \tiny{$\mu_{0,WVI}$} & \tiny{$\mu_{0,WVK}$} & \tiny{log$Z$ } & \tiny{
$\langle\mu_{0,W}\rangle$ } & \tiny{   log$Z$  } & \tiny{   $\langle\mu_{0,W}\rangle$  } & \tiny{ log$L/L_{can}$ } & \tiny{ log$Z$}\\
     &  & \tiny{ ($\pm$)      } & \tiny{ $\pm$0.06   } & \tiny{ $\pm$0.12   } & \tiny{ $\pm$0.10   } & \tiny{$\pm$0.15 } & \tiny{$\pm$0.15   } & \tiny{ $\pm$0.15}
 & \tiny{   $\pm$0.15  } & \tiny{ ($\pm$)} & \tiny{ $\pm$0.15    }\\
\tiny{1   } & \tiny{   2   } & \tiny{   3   } & \tiny{   4   } & \tiny{   5   } & \tiny{   6   } & \tiny{   7   } & \tiny{   8   } & \tiny{   9 } & \tiny{ 10 } & \tiny{ 11 } & \tiny{12   }\\
\hline
 \tiny{ BF Oph   } & \tiny{   0.609} & \tiny{   9.27(0.03)    } & \tiny{ 9.75} & \tiny{ 9.71} & \tiny{ 9.68} & \tiny{$-$1.84} & \tiny{ 9.70} & \tiny{$-$1.88} & \tiny{ 9.53} & \tiny{+0.50(0.15)} & \tiny{ $-$1.93 }\\
 \tiny{ T Vel    } & \tiny{   0.667} & \tiny{   9.80(0.06)    } & \tiny{10.47} & \tiny{10.41} & \tiny{10.31} & \tiny{$-$1.90} & \tiny{10.37} & \tiny{$-$1.94} & \tiny{10.20} & \tiny{+0.67(0.16)} & \tiny{$-$2.03 }\\
 \tiny{ $\delta$ Cep} & \tiny{0.730} & \tiny{   7.09(0.04)    } & \tiny{ 7.53} & \tiny{ 7.29} & \tiny{ 7.27} & \tiny{$-$2.08} & \tiny{ 7.31} & \tiny{$-$2.12} & \tiny{ 7.14} & \tiny{+0.26(0.16)} & \tiny{$-$2.13 }\\
 \tiny{ CV Mon   } & \tiny{   0.731} & \tiny{   10.99(0.03)   } & \tiny{11.77} & \tiny{11.26} & \tiny{11.32} & \tiny{$-$2.37} & \tiny{11.34} & \tiny{$-$2.41} & \tiny{11.17} & \tiny{+0.41(0.15)} & \tiny{$-$2.46 }\\
 \tiny{ V Cen    } & \tiny{   0.740} & \tiny{   9.18(0.06)    } & \tiny{ 9.73} & \tiny{ 9.47} & \tiny{ 9.43} & \tiny{$-$2.12} & \tiny{ 9.48} & \tiny{$-$2.16} & \tiny{ 9.30} & \tiny{+0.35(0.16)} & \tiny{$-$2.19 }\\
 \tiny{ BB Sgr   } & \tiny{   0.822} & \tiny{   9.52(0.03)    } & \tiny{ 9.77} & \tiny{ 9.65} & \tiny{ 9.70} & \tiny{$-$1.88} & \tiny{ 9.69} & \tiny{$-$1.92} & \tiny{ 9.52} & \tiny{+0.20(0.15)} & \tiny{$-$1.92 }\\
 \tiny{ U Sgr    } & \tiny{   0.829} & \tiny{   8.84(0.02)    } & \tiny{ 9.18} & \tiny{ 9.05} & \tiny{ 9.13} & \tiny{$-$1.89} & \tiny{ 9.10} & \tiny{$-$1.93} & \tiny{ 8.93} & \tiny{+0.31(0.15)} & \tiny{$-$1.95 }\\
 \tiny{ $\eta$ Aql  } & \tiny{0.856} & \tiny{   6.99(0.05)    } & \tiny{ 7.52} & \tiny{ 7.31} & \tiny{ 7.32} & \tiny{$-$2.03} & \tiny{ 7.34} & \tiny{$-$2.07} & \tiny{ 7.17} & \tiny{+0.41(0.16)} & \tiny{$-$2.11 }\\
 \tiny{ S Nor    } & \tiny{   0.989} & \tiny{   9.91(0.03)    } & \tiny{10.03} & \tiny{ 9.98} & \tiny{ 9.92} & \tiny{$-$1.88} & \tiny{ 9.96} & \tiny{$-$1.91} & \tiny{ 9.79} & \tiny{+0.06(0.15)} & \tiny{$-$1.89 }\\
 \tiny{ XX Cen   } & \tiny{   1.040} & \tiny{   11.11(0.02)   } & \tiny{11.49} & \tiny{11.33} & \tiny{11.32} & \tiny{$-$1.98} & \tiny{11.34} & \tiny{$-$2.01} & \tiny{11.17} & \tiny{+0.27(0.15)} & \tiny{$-$2.03 }\\
 \tiny{ V340 Nor } & \tiny{   1.053} & \tiny{   11.15(0.19)   } & \tiny{11.56} & \tiny{11.44} & \tiny{11.50} & \tiny{$-$1.88} & \tiny{11.48} & \tiny{$-$1.92} & \tiny{11.31} & \tiny{+0.39(0.24)} & \tiny{$-$1.96 }\\
 \tiny{ UU Mus   } & \tiny{   1.066} & \tiny{   12.59(0.08)   } & \tiny{13.03} & \tiny{12.83} & \tiny{12.77} & \tiny{$-$2.06} & \tiny{12.82} & \tiny{$-$2.10} & \tiny{12.65} & \tiny{+0.27(0.17)} & \tiny{$-$2.11 }\\
 \tiny{ U Nor    } & \tiny{   1.102} & \tiny{   10.72(0.06)   } & \tiny{11.14} & \tiny{10.92} & \tiny{10.91} & \tiny{$-$2.05} & \tiny{10.94} & \tiny{$-$2.09} & \tiny{10.77} & \tiny{+0.26(0.16)} & \tiny{$-$2.10 }\\
 \tiny{ BN Pup   } & \tiny{   1.136} & \tiny{   12.95(0.05)   } & \tiny{13.26} & \tiny{13.09} & \tiny{13.09} & \tiny{$-$1.98} & \tiny{13.11} & \tiny{$-$2.01} & \tiny{12.94} & \tiny{+0.18(0.16)} & \tiny{$-$2.01 }\\
 \tiny{ LS Pup   } & \tiny{   1.151} & \tiny{   13.55(0.06)   } & \tiny{13.71} & \tiny{13.59} & \tiny{13.57} & \tiny{$-$1.93} & \tiny{13.59} & \tiny{$-$1.97} & \tiny{13.42} & \tiny{+0.05(0.16)} & \tiny{$-$1.94 }\\
 \tiny{ VW Cen   } & \tiny{   1.177} & \tiny{   12.80(0.04)   } & \tiny{13.27} & \tiny{13.21} & \tiny{13.09} & \tiny{$-$1.93} & \tiny{13.16} & \tiny{$-$1.96} & \tiny{12.99} & \tiny{+0.42(0.16)} & \tiny{$-$2.01 }\\
 \tiny{ X Cyg    } & \tiny{   1.214} & \tiny{   10.42(0.02)   } & \tiny{10.15} & \tiny{10.29} & \tiny{10.31} & \tiny{$-$1.58} & \tiny{10.28} & \tiny{$-$1.62} & \tiny{10.11} & \tiny{$-$0.16(0.15)} & \tiny{$-$1.55 }\\
 \tiny{ VY Car   } & \tiny{   1.277} & \tiny{   11.50(0.02)   } & \tiny{11.45} & \tiny{11.52} & \tiny{11.48} & \tiny{$-$1.71} & \tiny{11.50} & \tiny{$-$1.74} & \tiny{11.33} & \tiny{$-$0.01(0.15)} & \tiny{$-$1.71  }\\
 \tiny{ RY Sco   } & \tiny{   1.308} & \tiny{   10.51(0.03)   } & \tiny{11.20} & \tiny{10.70} & \tiny{10.80} & \tiny{$-$2.34} & \tiny{10.80} & \tiny{$-$2.38} & \tiny{10.63} & \tiny{+0.33(0.15)} & \tiny{$-$2.41 }\\
 \tiny{ RZ Vel   } & \tiny{   1.310} & \tiny{   11.02(0.03)   } & \tiny{11.32} & \tiny{11.14} & \tiny{11.09} & \tiny{$-$2.03} & \tiny{11.13} & \tiny{$-$2.07} & \tiny{10.96} & \tiny{+0.13(0.15)} & \tiny{$-$2.06 }\\
 \tiny{ WZ Sgr   } & \tiny{   1.339} & \tiny{   11.29(0.05)   } & \tiny{11.49} & \tiny{11.40} & \tiny{11.37} & \tiny{$-$1.91} & \tiny{11.39} & \tiny{$-$1.94} & \tiny{11.22} & \tiny{+0.13(0.16)} & \tiny{$-$1.93 }\\
 \tiny{ WZ Car   } & \tiny{   1.362} & \tiny{   12.92(0.07)   } & \tiny{13.59} & \tiny{13.29} & \tiny{13.23} & \tiny{$-$2.19} & \tiny{13.30} & \tiny{$-$2.22} & \tiny{13.13} & \tiny{+0.44(0.17)} & \tiny{$-$2.27 }\\
 \tiny{ VZ Pup   } & \tiny{   1.365} & \tiny{   13.08(0.06)   } & \tiny{13.97} & \tiny{13.59} & \tiny{13.62} & \tiny{$-$2.22} & \tiny{13.65} & \tiny{$-$2.26} & \tiny{13.47} & \tiny{+0.65(0.16)} & \tiny{$-$2.35 }\\
 \tiny{ SW Vel   } & \tiny{   1.370} & \tiny{   12.00(0.03)   } & \tiny{12.49} & \tiny{12.23} & \tiny{12.19} & \tiny{$-$2.13} & \tiny{12.24} & \tiny{$-$2.17} & \tiny{12.07} & \tiny{+0.28(0.15)} & \tiny{$-$2.18 }\\
 \tiny{ T Mon    } & \tiny{   1.432} & \tiny{   10.82(0.06)   } & \tiny{10.67} & \tiny{10.78} & \tiny{10.72} & \tiny{$-$1.67} & \tiny{10.74} & \tiny{$-$1.71} & \tiny{10.57} & \tiny{$-$0.09(0.16)} & \tiny{$-$1.65 }\\
 \tiny{ RY Vel   } & \tiny{   1.449} & \tiny{   12.02(0.03)   } & \tiny{12.34} & \tiny{12.09} & \tiny{12.10} & \tiny{$-$2.08} & \tiny{12.12} & \tiny{$-$2.12} & \tiny{11.95} & \tiny{+0.11(0.15)        } & \tiny{$-$2.10 }\\
 \tiny{ AQ Pup   } & \tiny{   1.479} & \tiny{   12.52(0.05)   } & \tiny{12.76} & \tiny{12.45} & \tiny{12.56} & \tiny{$-$2.09} & \tiny{12.53} & \tiny{$-$2.13} & \tiny{12.36} & \tiny{+0.01(0.16)        } & \tiny{$-$2.09 }\\
 \tiny{ KN Cen   } & \tiny{   1.532} & \tiny{   13.12(0.05)   } & \tiny{13.34} & \tiny{13.03} & \tiny{12.84} & \tiny{$-$2.28} & \tiny{12.98} & \tiny{$-$2.32} & \tiny{12.81} & \tiny{$-$0.17(0.16)        } & \tiny{$-$2.24 }\\
 \tiny{ l Car    } & \tiny{   1.551} & \tiny{   8.99(0.03)    } & \tiny{ 8.39} & \tiny{ 8.70} & \tiny{ 8.69} & \tiny{$-$1.39} & \tiny{ 8.66} & \tiny{$-$1.42} & \tiny{ 8.49} & \tiny{$-$0.39(0.15)        } & \tiny{$-$1.32 }\\
 \tiny{ U Car    } & \tiny{   1.589} & \tiny{   10.97(0.03)   } & \tiny{11.44} & \tiny{11.24} & \tiny{11.22} & \tiny{$-$2.03} & \tiny{11.25} & \tiny{$-$2.07} & \tiny{11.08} & \tiny{+0.32(0.15)        } & \tiny{$-$2.09 }\\
 \tiny{ RS Pup   } & \tiny{   1.617} & \tiny{   11.56(0.06)   } & \tiny{11.48} & \tiny{11.31} & \tiny{11.36} & \tiny{$-$1.96} & \tiny{11.35} & \tiny{$-$1.99} & \tiny{11.18} & \tiny{ $-$0.25(0.16)       } & \tiny{$-$1.91 }\\
 \tiny{ SV Vul   } & \tiny{   1.653} & \tiny{   12.10(0.04)   } & \tiny{11.73} & \tiny{11.56} & \tiny{11.78} & \tiny{$-$1.85} & \tiny{11.68} & \tiny{$-$1.88} & \tiny{11.51} & \tiny{ $-$0.50(0.16)       } & \tiny{$-$1.75 }\\
\hline
\end{tabular}
\end{center}
\end{table*}
\linespread{1}

%table 9
\begin{table}\label{tab9}
\caption{Slopes of the $PL$ relations from canonical and
revised ISB distances to Galactic (MW) and LMC Cepheids, adopting
the correction suggested by G05. }
\begin{center}
\vspace{0.5truecm}
\begin{tabular}{lcccc}
\hline \hline
$M_i$ & MW & LMC & MW & LMC \\
      & (can) & (can) & (rev) & (rev) \\
\hline
$M_V$ & $-$3.08 &$-$3.05 &$-$2.90 &$-$2.87\\
$M_I$ & $-$3.30 &$-$3.29 &$-$3.13 &$-$3.11\\
$M_J$ & $-$3.53 &$-$3.48 &$-$3.33 &$-$3.29\\
$M_K$ & $-$3.67 &$-$3.54 &$-$3.46 &$-$3.36\\
\hline
\end{tabular}
\end{center}
\end{table}

%table 10
\linespread{0.7}
\begin{table*}\label{tab10}
\caption{Fundamental Galactic Cepheids with [Fe/H] measurements by
Andrievsky and collaborators. From left to right, the columns give:
name (1), period (2), reddening $E(B-V)$ (3), $BVI$ magnitudes
[(4)-(6)] and measured [Fe/H]$_A$ value (7). The pulsation metal
content log$Z$ and distance modulus $\mu_0$ at log$L/L_{can}$=0, as
inferred by the observed $WBV$ and $WVI$ functions, are listed in
columns (8) and (9), respectively, while the last two columns give
the resulting absolute magnitudes $M_V$ and $M_I$. }

\begin{center}
%\vspace{0.5truecm}
\begin{tabular}{lrrrrrrrrrr}
\hline \hline
\tiny{Name}&\tiny{log$P$}&\tiny{$E(B-V)$}&\tiny{$B$}&\tiny{$V$}&\tiny{$I$}&\tiny{[Fe/H]$_A$}&\tiny{log$Z$}&\tiny{$\mu_0$}&\tiny{$M_V$}&\tiny{$M_I$}\\
           &             &               &          &          &          &\tiny{$\pm$0.10} &\tiny{$\pm$0.15}& \tiny{$\pm$0.15}&\tiny{$\pm$0.15}&\tiny{$\pm$0.15} \\
\tiny{1} & \tiny{   2 } & \tiny{3 } & \tiny{4} & \tiny{5 } & \tiny{ 6        } & \tiny{   7      } & \tiny{   8      } & \tiny{  9      } & \tiny{ 10} & \tiny{ 11}   \\
\hline
\tiny{V504 Mon}&\tiny{0.443}&\tiny{0.57}&\tiny{12.82}&\tiny{11.81}&\tiny{10.54}&\tiny{$-$0.31}&\tiny{$-$2.06}&\tiny{12.91}&\tiny{$-$2.96}&\tiny{$-$3.49}\\
\tiny{FI Mon}   &   \tiny{  0.517   }   &   \tiny{  0.54}    &   \tiny{  14.07   }   &   \tiny{  12.95   }   &   \tiny{  11.57   }   &   \tiny{  $-$0.18 }   &   \tiny{  $-$2.01 }   &   \tiny{  14.01   }   &   \tiny{  $-$2.84 }   &   \tiny{  $-$3.51 }   \\
\tiny{V335 Aur}   &   \tiny{  0.533   }   &   \tiny{  0.66}    &   \tiny{  13.62   }   &   \tiny{  12.46   }   &   \tiny{  11.04   }   &   \tiny{  $-$0.27 }   &   \tiny{  $-$1.97 }   &   \tiny{  13.47   }   &   \tiny{  $-$3.18 }   &   \tiny{  $-$3.74 }   \\
\tiny{RT Aur}   &   \tiny{  0.571   }   &   \tiny{  0.05}    &   \tiny{  6.04    }   &   \tiny{  5.45    }   &   \tiny{  4.81    }   &   \tiny{+0.06    }   &   \tiny{  $-$1.87 }   &   \tiny{  8.55    }   &   \tiny{  $-$3.27 }   &   \tiny{  $-$3.84 }   \\
\tiny{SU Cyg}   &   \tiny{  0.585   }   &   \tiny{  0.10}    &   \tiny{  7.43    }   &   \tiny{  6.86    }   &   \tiny{  6.20    }   &   \tiny{  $-$0.01 }   &   \tiny{  $-$2.06 }   &   \tiny{  9.95    }   &   \tiny{  $-$3.40 }   &   \tiny{  $-$3.94 }   \\
\tiny{CS Ori}   &   \tiny{  0.590   }   &   \tiny{  0.41}    &   \tiny{  12.34   }   &   \tiny{  11.39   }   &   \tiny{  10.26   }   &   \tiny{  $-$0.26 }   &   \tiny{  $-$1.95 }   &   \tiny{  13.32   }   &   \tiny{  $-$3.28 }   &   \tiny{  $-$3.87 }   \\
\tiny{AE Tau}   &   \tiny{  0.591   }   &   \tiny{  0.60}    &   \tiny{  12.83   }   &   \tiny{  11.70   }   &   \tiny{  10.38   }   &   \tiny{  $-$0.19 }   &   \tiny{  $-$1.81 }   &   \tiny{  13.13   }   &   \tiny{  $-$3.42 }   &   \tiny{  $-$3.95 }   \\
\tiny{AA Mon}   &   \tiny{  0.595   }   &   \tiny{  0.83}    &   \tiny{  14.07   }   &   \tiny{  12.74   }   &   \tiny{  11.09   }   &   \tiny{  $-$0.21 }   &   \tiny{  $-$2.08 }   &   \tiny{  13.36   }   &   \tiny{  $-$3.37 }   &   \tiny{  $-$3.94 }   \\
\tiny{EK Mon}   &   \tiny{  0.597   }   &   \tiny{  0.58}    &   \tiny{  12.28   }   &   \tiny{  11.07   }   &   \tiny{  9.61    }   &   \tiny{  $-$0.10 }   &   \tiny{  $-$1.93 }   &   \tiny{  12.19   }   &   \tiny{  $-$3.05 }   &   \tiny{  $-$3.74 }   \\
\tiny{ST Tau}   &   \tiny{  0.606   }   &   \tiny{  0.36}    &   \tiny{  9.07    }   &   \tiny{  8.22    }   &   \tiny{  7.14    }   &   \tiny{  $-$0.05 }   &   \tiny{  $-$2.21 }   &   \tiny{  10.35   }   &   \tiny{  $-$3.31 }   &   \tiny{  $-$3.92 }   \\
\tiny{V495 Mon}   &   \tiny{  0.612   }   &   \tiny{  0.64}    &   \tiny{  13.69   }   &   \tiny{  12.44   }   &   \tiny{  10.97   }   &   \tiny{  $-$0.26 }   &   \tiny{  $-$1.80 }   &   \tiny{  13.57   }   &   \tiny{  $-$3.24 }   &   \tiny{  $-$3.87 }   \\
\tiny{V508 Mon}   &   \tiny{  0.616   }   &   \tiny{  0.32}    &   \tiny{  11.39   }   &   \tiny{  10.50   }   &   \tiny{  9.46    }   &   \tiny{  $-$0.25 }   &   \tiny{  $-$1.97 }   &   \tiny{  12.73   }   &   \tiny{  $-$3.30 }   &   \tiny{  $-$3.92 }   \\
\tiny{VW Pup}   &   \tiny{  0.632   }   &   \tiny{  0.51}    &   \tiny{  12.50   }   &   \tiny{  11.38   }   &   \tiny{  10.08   }   &   \tiny{  $-$0.19 }   &   \tiny{  $-$1.84 }   &   \tiny{  13.00   }   &   \tiny{  $-$3.32 }   &   \tiny{  $-$3.95 }   \\
\tiny{Y Lac}   &   \tiny{  0.636   }   &   \tiny{  0.22}    &   \tiny{  9.88    }   &   \tiny{  9.15    }   &   \tiny{  8.30    }   &   \tiny{  $-$0.09 }   &   \tiny{  $-$2.00 }   &   \tiny{  11.93   }   &   \tiny{  $-$3.50 }   &   \tiny{  $-$4.07 }   \\
\tiny{T Vul}   &   \tiny{  0.647   }   &   \tiny{  0.06}    &   \tiny{  6.40    }   &   \tiny{  5.75    }   &   \tiny{  5.08    }   &   \tiny{+0.01    }   &   \tiny{  $-$1.83 }   &   \tiny{  9.00    }   &   \tiny{  $-$3.45 }   &   \tiny{  $-$4.05 }   \\
\tiny{FF Aql}   &   \tiny{  0.650   }   &   \tiny{  0.22}    &   \tiny{  6.13    }   &   \tiny{  5.37    }   &   \tiny{  4.51    }   &   \tiny{+0.02    }   &   \tiny{  $-$1.97 }   &   \tiny{  8.16    }   &   \tiny{  $-$3.53 }   &   \tiny{  $-$4.10 }   \\
\tiny{FG Mon}   &   \tiny{  0.653   }   &   \tiny{  0.68}    &   \tiny{  14.44   }   &   \tiny{  13.24   }   &   \tiny{  11.74   }   &   \tiny{  $-$0.20 }   &   \tiny{  $-$2.10 }   &   \tiny{  14.47   }   &   \tiny{  $-$3.48 }   &   \tiny{  $-$4.08 }   \\
\tiny{T Vel}   &   \tiny{  0.667   }   &   \tiny{  0.28}    &   \tiny{  8.97    }   &   \tiny{  8.04    }   &   \tiny{  6.96    }   &   \tiny{  $-$0.02 }   &   \tiny{  $-$1.90 }   &   \tiny{  10.34   }   &   \tiny{  $-$3.24 }   &   \tiny{  $-$3.94 }   \\
\tiny{WW Mon}   &   \tiny{  0.669   }   &   \tiny{  0.64}    &   \tiny{  13.65   }   &   \tiny{  12.50   }   &   \tiny{  11.14   }   &   \tiny{  $-$0.29 }   &   \tiny{  $-$1.95 }   &   \tiny{  14.09   }   &   \tiny{  $-$3.68 }   &   \tiny{  $-$4.22 }   \\
\tiny{RY CMa}   &   \tiny{  0.670   }   &   \tiny{  0.25 }   &   \tiny{  8.95    }   &   \tiny{  8.11    }   &   \tiny{  7.13    }   &   \tiny{+0.02    }   &   \tiny{  $-$1.94 }   &   \tiny{  10.69   }   &   \tiny{  $-$3.40 }   &   \tiny{  $-$4.05 }   \\
\tiny{CU Mon}   &   \tiny{  0.673   }   &   \tiny{  0.79}    &   \tiny{  15.02   }   &   \tiny{  13.61   }   &   \tiny{  11.90   }   &   \tiny{  $-$0.26 }   &   \tiny{  $-$1.94 }   &   \tiny{  14.36   }   &   \tiny{  $-$3.35 }   &   \tiny{  $-$4.02 }   \\
\tiny{EE Mon}   &   \tiny{  0.682   }   &   \tiny{  0.49}    &   \tiny{  14.03   }   &   \tiny{  13.02   }   &   \tiny{  11.70   }   &   \tiny{  $-$0.51 }   &   \tiny{  $-$2.35 }   &   \tiny{  14.81   }   &   \tiny{  $-$3.41 }   &   \tiny{  $-$4.08 }   \\
\tiny{CF Cas}   &   \tiny{  0.688   }   &   \tiny{  0.57}    &   \tiny{  12.34   }   &   \tiny{  11.14   }   &   \tiny{  9.75    }   &   \tiny{  $-$0.01 }   &   \tiny{  $-$1.78 }   &   \tiny{  12.74   }   &   \tiny{  $-$3.47 }   &   \tiny{  $-$4.11 }   \\
\tiny{BG Lac}   &   \tiny{  0.727   }   &   \tiny{  0.34}    &   \tiny{  9.85    }   &   \tiny{  8.89    }   &   \tiny{  7.81    }   &   \tiny{  $-$0.01 }   &   \tiny{  $-$1.83 }   &   \tiny{  11.40   }   &   \tiny{  $-$3.62 }   &   \tiny{  $-$4.25 }   \\
\tiny{$\delta$ Cep}   &   \tiny{  0.730   }   &   \tiny{  0.09}    &   \tiny{  4.62    }   &   \tiny{  3.96    }   &   \tiny{  3.20    }   &   \tiny{+0.06    }   &   \tiny{  $-$2.06 }   &   \tiny{  7.30    }   &   \tiny{  $-$3.65 }   &   \tiny{  $-$4.28 }   \\
\tiny{CV Mon}   &   \tiny{  0.731   }   &   \tiny{  0.71}    &   \tiny{  11.61   }   &   \tiny{  10.30   }   &   \tiny{  8.65    }   &   \tiny{  $-$0.03 }   &   \tiny{  $-$2.26 }   &   \tiny{  11.39   }   &   \tiny{  $-$3.44 }   &   \tiny{  $-$4.16 }   \\
\tiny{XX Mon}   &   \tiny{  0.737   }   &   \tiny{  0.60}    &   \tiny{  13.10   }   &   \tiny{  11.92   }   &   \tiny{  10.50   }   &   \tiny{  $-$0.10 }   &   \tiny{  $-$1.98 }   &   \tiny{  13.61   }   &   \tiny{  $-$3.66 }   &   \tiny{  $-$4.30 }   \\
\tiny{V Cen}   &   \tiny{  0.740   }   &   \tiny{  0.29}    &   \tiny{  7.69    }   &   \tiny{  6.82    }   &   \tiny{  5.81    }   &   \tiny{+0.04    }   &   \tiny{  $-$2.00 }   &   \tiny{  9.53    }   &   \tiny{  $-$3.67 }   &   \tiny{  $-$4.30 }   \\
\tiny{WW Pup}   &   \tiny{  0.742   }   &   \tiny{  0.40}    &   \tiny{  11.47   }   &   \tiny{  10.61   }   &   \tiny{  9.51    }   &   \tiny{  $-$0.18 }   &   \tiny{  $-$2.39 }   &   \tiny{  13.13   }   &   \tiny{  $-$3.83 }   &   \tiny{  $-$4.42 }   \\
\tiny{RZ Gem}   &   \tiny{  0.743   }   &   \tiny{  0.57}    &   \tiny{  11.06   }   &   \tiny{  10.02   }   &   \tiny{  8.75    }   &   \tiny{  $-$0.12 }   &   \tiny{  $-$2.16 }   &   \tiny{  12.10   }   &   \tiny{  $-$3.96 }   &   \tiny{  $-$4.49 }   \\
\tiny{Y Sgr}   &   \tiny{  0.761   }   &   \tiny{  0.21}    &   \tiny{  6.60    }   &   \tiny{  5.74    }   &   \tiny{  4.78    }   &   \tiny{+0.06    }   &   \tiny{  $-$1.94 }   &   \tiny{  8.65    }   &   \tiny{  $-$3.58 }   &   \tiny{  $-$4.28 }   \\
\tiny{FM Aql}   &   \tiny{  0.786   }   &   \tiny{  0.65}    &   \tiny{  9.58    }   &   \tiny{  8.28    }   &   \tiny{  6.78    }   &   \tiny{+0.08    }   &   \tiny{  $-$1.76 }   &   \tiny{  9.92    }   &   \tiny{  $-$3.78 }   &   \tiny{  $-$4.43 }   \\
\tiny{X Vul}   &   \tiny{  0.801   }   &   \tiny{  0.85}    &   \tiny{  10.24   }   &   \tiny{  8.84    }   &   \tiny{  7.20    }   &   \tiny{+0.08    }   &   \tiny{  $-$1.88 }   &   \tiny{  10.16   }   &   \tiny{  $-$4.11 }   &   \tiny{  $-$4.65 }   \\
\tiny{U Sgr}   &   \tiny{  0.829   }   &   \tiny{  0.40}    &   \tiny{  7.79    }   &   \tiny{  6.70    }   &   \tiny{  5.45    }   &   \tiny{+0.04    }   &   \tiny{  $-$1.87 }   &   \tiny{  9.11    }   &   \tiny{  $-$3.74 }   &   \tiny{  $-$4.46 }   \\
\tiny{TW Mon}   &   \tiny{  0.851   }   &   \tiny{  0.70}    &   \tiny{  13.91   }   &   \tiny{  12.57   }   &   \tiny{  10.95   }   &   \tiny{  $-$0.24 }   &   \tiny{  $-$2.01 }   &   \tiny{  14.14   }   &   \tiny{  $-$3.87 }   &   \tiny{  $-$4.57 }   \\
\tiny{$\eta$ Aql}   &   \tiny{  0.856   }   &   \tiny{  0.15}    &   \tiny{  4.69    }   &   \tiny{  3.90    }   &   \tiny{  3.03    }   &   \tiny{+0.05    }   &   \tiny{  $-$1.95 }   &   \tiny{  7.37    }   &   \tiny{  $-$3.97 }   &   \tiny{  $-$4.63 }   \\
\tiny{V510 Mon}   &   \tiny{  0.864   }   &   \tiny{  0.84}    &   \tiny{  14.13   }   &   \tiny{  12.65   }   &   \tiny{  10.86   }   &   \tiny{  $-$0.19 }   &   \tiny{  $-$2.00 }   &   \tiny{  13.82   }   &   \tiny{  $-$3.95 }   &   \tiny{  $-$4.64 }   \\
\tiny{TZ Mon}   &   \tiny{  0.871   }   &   \tiny{  0.44}    &   \tiny{  11.92   }   &   \tiny{  10.79   }   &   \tiny{  9.47    }   &   \tiny{  $-$0.12 }   &   \tiny{  $-$1.99 }   &   \tiny{  13.16   }   &   \tiny{  $-$3.83 }   &   \tiny{  $-$4.57 }   \\
\tiny{W Sgr}   &   \tiny{  0.881   }   &   \tiny{  0.11}    &   \tiny{  5.41    }   &   \tiny{  4.67    }   &   \tiny{  3.85    }   &   \tiny{  $-$0.01 }   &   \tiny{  $-$2.03 }   &   \tiny{  8.33    }   &   \tiny{  $-$4.03 }   &   \tiny{  $-$4.71 }   \\
\tiny{RX Cam}   &   \tiny{  0.898   }   &   \tiny{  0.57}    &   \tiny{  8.88    }   &   \tiny{  7.68    }   &   \tiny{  6.26    }   &   \tiny{+0.03    }   &   \tiny{  $-$2.05 }   &   \tiny{  9.89    }   &   \tiny{  $-$4.09 }   &   \tiny{  $-$4.77 }   \\
\tiny{W Gem}   &   \tiny{  0.898   }   &   \tiny{  0.28}    &   \tiny{  7.87    }   &   \tiny{  6.95    }   &   \tiny{  5.97    }   &   \tiny{  $-$0.04 }   &   \tiny{  $-$1.84 }   &   \tiny{  10.26   }   &   \tiny{  $-$4.24 }   &   \tiny{  $-$4.85 }   \\
\tiny{U Vul}   &   \tiny{  0.903   }   &   \tiny{  0.65}    &   \tiny{  8.41    }   &   \tiny{  7.13    }   &   \tiny{  5.61    }   &   \tiny{+0.05    }   &   \tiny{  $-$2.05 }   &   \tiny{  9.11    }   &   \tiny{  $-$4.14 }   &   \tiny{  $-$4.80 }   \\
\tiny{DL Cas}   &   \tiny{  0.903   }   &   \tiny{  0.53}    &   \tiny{  10.12   }   &   \tiny{  8.97    }   &   \tiny{  7.66    }   &   \tiny{  $-$0.01 }   &   \tiny{  $-$1.92 }   &   \tiny{  11.46   }   &   \tiny{  $-$4.25 }   &   \tiny{  $-$4.87 }   \\
\tiny{AC Mon}   &   \tiny{  0.904   }   &   \tiny{  0.51}    &   \tiny{  11.28   }   &   \tiny{  10.10   }   &   \tiny{  8.71    }   &   \tiny{  $-$0.22 }   &   \tiny{  $-$2.04 }   &   \tiny{  12.41   }   &   \tiny{  $-$3.99 }   &   \tiny{  $-$4.72 }   \\
\tiny{S Sge}   &   \tiny{  0.923   }   &   \tiny{  0.13}    &   \tiny{  6.42    }   &   \tiny{  5.61    }   &   \tiny{  4.78    }   &   \tiny{+0.10    }   &   \tiny{  $-$1.84 }   &   \tiny{  9.37    }   &   \tiny{  $-$4.18 }   &   \tiny{  $-$4.85 }   \\
\tiny{TX Mon}   &   \tiny{  0.940   }   &   \tiny{  0.51}    &   \tiny{  12.08   }   &   \tiny{  10.97   }   &   \tiny{  9.63    }   &   \tiny{  $-$0.14 }   &   \tiny{  $-$2.16 }   &   \tiny{  13.55   }   &   \tiny{  $-$4.27 }   &   \tiny{  $-$4.93 }   \\
\tiny{FN Aql}   &   \tiny{  0.977   }   &   \tiny{  0.51}    &   \tiny{  9.62    }   &   \tiny{  8.38    }   &   \tiny{  7.00    }   &   \tiny{  $-$0.02 }   &   \tiny{  $-$1.80 }   &   \tiny{  10.93   }   &   \tiny{  $-$4.24 }   &   \tiny{  $-$4.95 }   \\
\tiny{SX Vel}   &   \tiny{  0.980   }   &   \tiny{  0.25}    &   \tiny{  9.17    }   &   \tiny{  8.26    }   &   \tiny{  7.25    }   &   \tiny{  $-$0.03 }   &   \tiny{  $-$2.02 }   &   \tiny{  11.78   }   &   \tiny{  $-$4.34 }   &   \tiny{  $-$5.02 }   \\
\tiny{YZ Sgr}   &   \tiny{  0.980   }   &   \tiny{  0.29}    &   \tiny{  8.38    }   &   \tiny{  7.34    }   &   \tiny{  6.21    }   &   \tiny{+0.05    }   &   \tiny{  $-$1.86 }   &   \tiny{  10.55   }   &   \tiny{  $-$4.18 }   &   \tiny{  $-$4.92 }   \\
\tiny{S Nor}   &   \tiny{  0.989   }   &   \tiny{  0.19}    &   \tiny{  7.37    }   &   \tiny{  6.43    }   &   \tiny{  5.42    }   &   \tiny{+0.05    }   &   \tiny{  $-$1.86 }   &   \tiny{  9.97    }   &   \tiny{  $-$4.17 }   &   \tiny{  $-$4.93 }   \\
\tiny{$\beta$ Dor}   &   \tiny{  0.993   }   &   \tiny{  0.04}    &   \tiny{  4.56    }   &   \tiny{  3.75    }   &   \tiny{  2.94    }   &   \tiny{  $-$0.01 }   &   \tiny{  $-$1.83 }   &   \tiny{  7.80    }   &   \tiny{  $-$4.19 }   &   \tiny{  $-$4.95 }   \\
\tiny{$\zeta$ Gem}   &   \tiny{  1.006   }   &   \tiny{  0.02}    &   \tiny{  4.71    }   &   \tiny{  3.90    }   &   \tiny{  3.10    }   &   \tiny{+0.04    }   &   \tiny{  $-$1.75 }   &   \tiny{  8.01    }   &   \tiny{  $-$4.17 }   &   \tiny{  $-$4.95 }   \\
\tiny{Z Lac}   &   \tiny{  1.037   }   &   \tiny{  0.40}    &   \tiny{  9.51    }   &   \tiny{  8.42    }   &   \tiny{  7.20    }   &   \tiny{+0.01    }   &   \tiny{  $-$1.93 }   &   \tiny{  11.59   }   &   \tiny{  $-$4.51 }   &   \tiny{  $-$5.20 }   \\
\tiny{VX Per}   &   \tiny{  1.037   }   &   \tiny{  0.52}    &   \tiny{  10.46   }   &   \tiny{  9.31    }   &   \tiny{  8.00    }   &   \tiny{  $-$0.05 }   &   \tiny{  $-$1.98 }   &   \tiny{  12.25   }   &   \tiny{  $-$4.65 }   &   \tiny{  $-$5.28 }   \\
\tiny{AA Gem}   &   \tiny{  1.053   }   &   \tiny{  0.33}    &   \tiny{  10.80   }   &   \tiny{  9.73    }   &   \tiny{  8.58    }   &   \tiny{  $-$0.24 }   &   \tiny{  $-$1.81 }   &   \tiny{  13.13   }   &   \tiny{  $-$4.49 }   &   \tiny{  $-$5.20 }   \\
\tiny{RX Aur}   &   \tiny{  1.065   }   &   \tiny{  0.28}    &   \tiny{  8.63    }   &   \tiny{  7.67    }   &   \tiny{  6.66    }   &   \tiny{  $-$0.07 }   &   \tiny{  $-$1.89 }   &   \tiny{  11.46   }   &   \tiny{  $-$4.69 }   &   \tiny{  $-$5.35 }   \\
\tiny{HW Pup}   &   \tiny{  1.129   }   &   \tiny{  0.72}    &   \tiny{  13.35   }   &   \tiny{  12.10   }   &   \tiny{  10.55   }   &   \tiny{  $-$0.20 }   &   \tiny{  $-$2.37 }   &   \tiny{  14.78   }   &   \tiny{  $-$5.07 }   &   \tiny{  $-$5.67 }   \\
\tiny{VY Sgr}   &   \tiny{  1.132   }   &   \tiny{  1.28}    &   \tiny{  13.46   }   &   \tiny{  11.45   }   &   \tiny{  9.19    }   &   \tiny{+0.26    }   &   \tiny{  $-$1.50 }   &   \tiny{  12.28   }   &   \tiny{  $-$5.07 }   &   \tiny{  $-$5.65 }   \\
\tiny{AD Pup}   &   \tiny{  1.133   }   &   \tiny{  0.33}    &   \tiny{  10.94   }   &   \tiny{  9.90    }   &   \tiny{  8.71    }   &   \tiny{  $-$0.24 }   &   \tiny{  $-$2.12 }   &   \tiny{  13.49   }   &   \tiny{  $-$4.68 }   &   \tiny{  $-$5.43 }   \\
\tiny{BN Pup}   &   \tiny{  1.136   }   &   \tiny{  0.44}    &   \tiny{  11.09   }   &   \tiny{  9.89    }   &   \tiny{  8.56    }   &   \tiny{+0.01    }   &   \tiny{  $-$1.95 }   &   \tiny{  13.11   }   &   \tiny{  $-$4.66 }   &   \tiny{  $-$5.42 }   \\
\tiny{TT Aql}   &   \tiny{  1.138   }   &   \tiny{  0.50}    &   \tiny{  8.43    }   &   \tiny{  7.13    }   &   \tiny{  5.72    }   &   \tiny{+0.11    }   &   \tiny{  $-$1.73 }   &   \tiny{  10.15   }   &   \tiny{  $-$4.65 }   &   \tiny{  $-$5.41 }   \\
\tiny{UZ Sct}   &   \tiny{  1.169   }   &   \tiny{  1.07}    &   \tiny{  13.15   }   &   \tiny{  11.25   }   &   \tiny{  9.18    }   &   \tiny{+0.33    }   &   \tiny{  $-$1.37 }   &   \tiny{  12.68   }   &   \tiny{  $-$4.96 }   &   \tiny{  $-$5.63 }   \\
\tiny{SV Mon}   &   \tiny{  1.183   }   &   \tiny{  0.25}    &   \tiny{  9.31    }   &   \tiny{  8.27    }   &   \tiny{  7.13    }   &   \tiny{  0.00    }   &   \tiny{  $-$1.97 }   &   \tiny{  12.14   }   &   \tiny{  $-$4.69 }   &   \tiny{  $-$5.50 }   \\
\tiny{AV Sgr}   &   \tiny{  1.188   }   &   \tiny{  1.27}    &   \tiny{  13.39   }   &   \tiny{  11.30   }   &   \tiny{  8.88    }   &   \tiny{+0.34    }   &   \tiny{  $-$1.67 }   &   \tiny{  11.94   }   &   \tiny{  $-$4.83 }   &   \tiny{  $-$5.58 }   \\
\tiny{X Cyg}   &   \tiny{  1.214   }   &   \tiny{  0.29}    &   \tiny{  7.53    }   &   \tiny{  6.39    }   &   \tiny{  5.24    }   &   \tiny{+0.12    }   &   \tiny{  $-$1.68 }   &   \tiny{  10.29   }   &   \tiny{  $-$4.85 }   &   \tiny{  $-$5.63 }   \\
\tiny{CD Cyg}   &   \tiny{  1.232   }   &   \tiny{  0.51}    &   \tiny{  10.25   }   &   \tiny{  8.95    }   &   \tiny{  7.50    }   &   \tiny{+0.07    }   &   \tiny{  $-$1.93 }   &   \tiny{  12.19   }   &   \tiny{  $-$4.94 }   &   \tiny{  $-$5.71 }   \\
\tiny{SZ Aql}   &   \tiny{  1.234   }   &   \tiny{  0.64}    &   \tiny{  10.06   }   &   \tiny{  8.63    }   &   \tiny{  7.07    }   &   \tiny{+0.15    }   &   \tiny{  $-$1.76 }   &   \tiny{  11.57   }   &   \tiny{  $-$5.05 }   &   \tiny{  $-$5.78 }   \\
\tiny{RZ Vel}   &   \tiny{  1.310   }   &   \tiny{  0.34}    &   \tiny{  8.21    }   &   \tiny{  7.08    }   &   \tiny{  5.86    }   &   \tiny{  $-$0.07 }   &   \tiny{  $-$2.01 }   &   \tiny{  11.14   }   &   \tiny{  $-$5.17 }   &   \tiny{  $-$5.96 }   \\
\tiny{V340 Ara}   &   \tiny{  1.318   }   &   \tiny{  0.57}    &   \tiny{  11.76   }   &   \tiny{  10.21   }   &   \tiny{  8.58    }   &   \tiny{+0.31    }   &   \tiny{  $-$1.52 }   &   \tiny{  13.23   }   &   \tiny{  $-$4.92 }   &   \tiny{  $-$5.80 }   \\
\tiny{WZ Sgr}   &   \tiny{  1.339   }   &   \tiny{  0.47}    &   \tiny{  9.43    }   &   \tiny{  8.03    }   &   \tiny{  6.53    }   &   \tiny{+0.17    }   &   \tiny{  $-$1.76 }   &   \tiny{  11.48   }   &   \tiny{  $-$4.99 }   &   \tiny{  $-$5.88 }   \\
\tiny{VZ Pup}   &   \tiny{  1.365   }   &   \tiny{  0.47}   &   \tiny{  10.78   }   &   \tiny{  9.63    }   &   \tiny{  8.30    }   &   \tiny{  $-$0.16 }   &   \tiny{  $-$2.25 }   &   \tiny{  13.64   }   &   \tiny{  $-$5.57 }   &   \tiny{  $-$6.28 }   \\
\tiny{SW Vel}   &   \tiny{  1.370   }   &   \tiny{  0.35 }   &   \tiny{  9.27    }   &   \tiny{  8.12    }   &   \tiny{  6.84    }   &   \tiny{  $-$0.07 }   &   \tiny{  $-$2.13 }   &   \tiny{  12.24   }   &   \tiny{  $-$5.28 }   &   \tiny{  $-$6.10 }   \\
\tiny{X Pup}   &   \tiny{  1.414   }   &   \tiny{  0.44}    &   \tiny{  9.73    }   &   \tiny{  8.53    }   &   \tiny{  7.17    }   &   \tiny{  $-$0.03 }   &   \tiny{  $-$2.16 }   &   \tiny{  12.62   }   &   \tiny{  $-$5.56 }   &   \tiny{  $-$6.33 }   \\
\tiny{T Mon}   &   \tiny{  1.432   }   &   \tiny{  0.21}    &   \tiny{  7.29    }   &   \tiny{  6.13    }   &   \tiny{  4.98    }   &   \tiny{+0.13    }   &   \tiny{  $-$1.67 }   &   \tiny{  10.77   }   &   \tiny{  $-$5.33 }   &   \tiny{  $-$6.20 }   \\
\tiny{RY Vel}   &   \tiny{  1.449   }   &   \tiny{  0.56}    &   \tiny{  9.74    }   &   \tiny{  8.37    }   &   \tiny{  6.83    }   &   \tiny{  $-$0.03 }   &   \tiny{  $-$2.14 }   &   \tiny{  12.10   }   &   \tiny{  $-$5.58 }   &   \tiny{  $-$6.39 }   \\
\tiny{KQ Sco}   &   \tiny{  1.458   }   &   \tiny{  0.90}    &   \tiny{  11.75   }   &   \tiny{  9.81    }   &   \tiny{  7.66    }   &   \tiny{+0.16    }   &   \tiny{  $-$1.68 }   &   \tiny{  12.00   }   &   \tiny{  $-$5.15 }   &   \tiny{  $-$6.13 }   \\
\tiny{AQ Pup}   &   \tiny{  1.479   }   &   \tiny{  0.51}    &   \tiny{  10.07   }   &   \tiny{  8.70    }   &   \tiny{  7.14    }   &   \tiny{  $-$0.14 }   &   \tiny{  $-$2.20 }   &   \tiny{  12.50   }   &   \tiny{  $-$5.48 }   &   \tiny{  $-$6.37 }   \\
\tiny{RS Pup}   &   \tiny{  1.617   }   &   \tiny{  0.45 }   &   \tiny{  8.46    }   &   \tiny{  7.03    }   &   \tiny{  5.49    }   &   \tiny{+0.17    }   &   \tiny{  $-$1.98 }   &   \tiny{  11.31   }   &   \tiny{  $-$5.74 }   &   \tiny{  $-$6.70 }   \\
\tiny{SV Vul}   &   \tiny{  1.653   }   &   \tiny{  0.57}    &   \tiny{  8.67    }   &   \tiny{  7.21    }   &   \tiny{  5.69    }   &   \tiny{+0.03    }   &   \tiny{  $-$1.78 }   &   \tiny{  11.65   }   &   \tiny{  $-$6.32 }   &   \tiny{  $-$7.09 }   \\
\tiny{S Vul}   &   \tiny{  1.835   }   &   \tiny{  0.83}    &   \tiny{  10.85   }   &   \tiny{  8.96    }   &   \tiny{  6.94    }   &   \tiny{  $-$0.02 }   &   \tiny{  $-$1.72 }   &   \tiny{  12.73   }   &   \tiny{  $-$6.50 }   &   \tiny{  $-$7.44 }   \\

\hline
\end{tabular}
\end{center}
\end{table*}
\linespread{1}

\subsection{Galactic Cepheids with measured [Fe/H]}

The evolutionary $PW_e$ relations presented in Table 3 give the
distance to Cepheids with known metal content, once the
log$L/L_{can}$ ratio is adopted. On the other hand, the opposite
metallicity dependence of $WBV$ with respect to $WVI$ and $WVK$
provides a direct way to estimate the metal content of Cepheids
with $BVI$ or $BVK$ magnitudes, independently of distance and
reddening and scarcely influenced by the $L/L_{can}$ ratio.

%\fig11
\begin{figure}
\includegraphics[width=8cm]{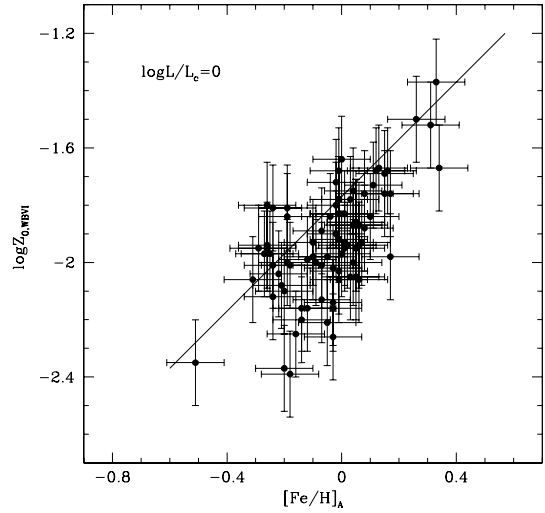}
\caption{\small{Pulsation metal content of Galactic Cepheids
determined by the predicted $P$-$WBV$ and $P$-$WVI$ relations versus
the spectroscopic measurement [Fe/H]$_A$ by Andrievski and
coworkers. The solid line is the relation log$Z$=[Fe/H]$-$1.77 with
$Z_{\odot}$=0.017.}} \label{7587fig11}
\end{figure}

%\fig12
\begin{figure}
\includegraphics[width=8cm]{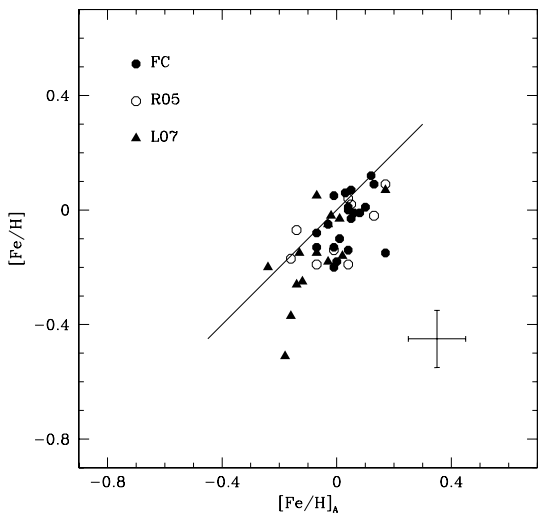}
\caption{\small{Spectroscopic measurements by Fry \& Carney
(1997:FC), Romaniello et al. (2005: R05) and Lemasle et al. (2007:
L07) in comparison with the Andrievky and coworkers values. The solid line
is the equality line and the large cross
shows the average uncertainty of the spectroscopic data.}}
\label{7587fig12}
\end{figure}

In order to test the reliability of the method, we use the
observed $BVI$ magnitudes compiled by Berdnikov, Dambis \&
Vozyakova (2000) for fundamental pulsators together with the
extensive [Fe/H] measurements by Andrievsky and coworkers
(Andrievsky 2002a,b,c, 2004; Luck et al. 2003). The pulsation
log$Z$ and $\mu_0$ values at log$L/L_{can}$=0, as derived by the
observed $WBV$ and $WVI$ quantities together with the predicted
relations given in Table 3, are listed in columns (8) and (9),
respectively, in Table 10, while the last two columns give the
resulting absolute $M_V$ and $M_I$ magnitudes adopting the
$E(B-V)$ reddening (Fernie et al. 1995) given in column (3) in the
Table.

Figure 11 shows the comparison between the measured [Fe/H]$_A$
parameter and the pulsation metal content log$Z$ at
log$L/L_{can}$=0. {\bf We find log$Z$=[Fe/H]$_A-$1.90$\pm$0.15 which
is somehow different from the canonical relation
log$Z$=[Fe/H]$-$1.77 (solid line) with $Z_{\odot}$=0.017.} However,
given the current discrepancy {\bf (see also Groenewegen et al. 2004)}
between abundance determinations by
different authors, as shown in Fig. 12, the pulsation method seems
to be sufficiently adequate to provide the Cepheid metal content
with an uncertainty almost comparable with that of spectroscopic
studies.

%\fig13
\begin{figure}
\includegraphics[width=8cm]{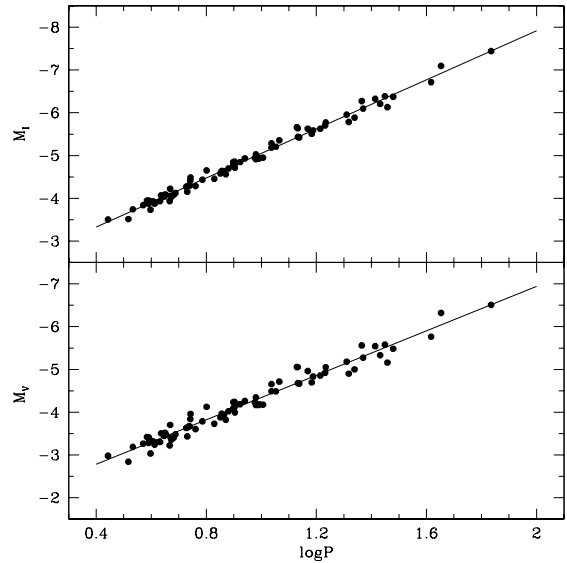}
\caption{\small{$M_V$ and $M_I$ magnitudes versus period for
Galactic Cepheids with spectroscopic [Fe/H]$_A$ measurements. The
solid lines are drawn by a linear regression through the points.}} \label{7587fig13}
\end{figure}

{\bf Using the pulsation distances in column (9) and the reddening
in column (3), we determine the Cepheid absolute magnitudes $M_V$
and $M_I$ listed in the last two columns in Table 10. As shown in
Fig. 13, the resulting $P$-$M_V$ and $P$-$M_I$ relations of these
Galactic Cepheids with [Fe/H]=$-0.04\pm$0.16 are well defined and
the linear regression through the points (solid line) is
$$M_V=-1.74(\pm0.15)-2.60(\pm0.17)\log P\eqno(7)$$
\noindent and
$$M_I=-2.18(\pm0.09)-2.87(\pm0.10)\log P\eqno(8)$$
\noindent whose slopes are closer to the B07 than to the S04
results and {\it shallower} than the LMC observed values listed in
Table 11. Furthermore, we find that for the Cepheids with
[Fe/H]$\le$0 the slope of the $P$-$M_V$ and $P$-$M_I$ relations is
$-$2.69 and $-$2.94, respectively, whereas for those with
[Fe/H]$\ge$0 we get $-$2.49 and $-$2.80, even more supporting the
predicted dependence of the optical $PL$ relations slope on the
metal content. We wish to emphasize the reliability of these
results because they are independent of the log$L/L_{can}$ ratio,
which influences only the zero-point, and are based on the
predicted evolutionary $P$-$WBV$ and $P$-$WVI$ relations whose
slopes are the {\it same} as the LMC observed relations. We will
come back to this issue at the end of this section.}

\subsection{Magellanic Cloud Cepheids}

The $BVI$ magnitudes of the many hundreds of LMC Cepheids
collected during the OGLE-II micro-lensing survey (Udalski et al.
1999) and the near-infrared $JK_s$ measurements recently presented
by P04 for 92 variables yield the apparent $PL$ and $PW$ relations
listed in Table 11. The slopes of the observed $PL$ relations appear
quite consistent with those predicted by the synthetic linear $PL$
relations at $Z$=0.008, the canonical metal content of LMC
variables, and compared with those provided by the B07 Galactic
Cepheids support the predicted steepening of the $PL$ relations
when moving from metal-rich to metal-poor variables. {\bf
Regarding the $PW$ relations, the observed slopes are also in close
agreement with the predicted values listed in Table 3, {\it
excluding a significant variation of log$L/L_{can}$ with the
period among the LMC Cepheids.}

%table 11
\begin{table}\label{tab11}
\caption{Apparent $PL$ and $PW$ relations for LMC Cepheids,
as based on $VI$ magnitudes (Udalski et al. 1999) and 2Mass
near-infrared magnitudes (P04).}
\begin{center}
\vspace{0.5truecm}
\begin{tabular}{lccclccc}
\hline \hline
$M_i$ & $\alpha$ & $\beta$ &$\sigma$ & $W$ & $\alpha$ & $\beta$ &$\sigma$ \\
\hline
\\
$B_0$   &  17.37 & $-$2.44 &0.24  & $WBV$   & 16.01 & $-$3.83&0.22\\
$V_0$   &  17.07 & $-$2.78 &0.16  & $WVI$   & 15.88 & $-$3.29&0.08\\
$I_0$   &  16.59 & $-$2.98 &0.11  &         &       &        &\\
$J_0$   &  16.34 & $-$3.15 &      &         &       &        &\\
$K_{s0}$&  16.04 & $-$3.26 &      & $WJK_s$ & 15.90 & $-$3.36&0.05\\
\hline
\end{tabular}
\end{center}
\end{table}

By comparison of the predicted evolutionary $PW$ relations with the
observed $WBV$, $WVI$ and $WJK$\footnote{The 2Mass $JK_s$ magnitudes
given by P04 are transformed into the Bessel \&
Brett system adopting the relations given by Carpenter (2001).}
functions, we derive
$$\mu_{0,WBV}=20.21(\pm0.20)-0.72\log L/L_{can}+0.71\log Z$$
$$\mu_{0,WVI}=18.56(\pm0.08)-0.84\log L/L_{can}-0.08\log Z$$
$$\mu_{0,WJK}=18.57(\pm0.08)-0.90\log L/L_{can}-0.08\log Z$$

\noindent where the large uncertainty of the $WBV$-based distance
reflects the observed dispersion around the $P$-$WBV$ relation which
is likely due to less accurate sampling of the light curves and/or
to a moderate metallicity spread among LMC Cepheids. In any case, our
pulsation approach yields that
the agreement among the $PW$-based distances is achieved at
$Z\sim$0.008, which is the classical metal content of LMC Cepheids.
Using only the $WVI$ and $WJK$
functions which are mildly metallicity dependent, we eventually find
$\mu_0$(LMC)=18.52$\pm$0.11 mag at $Z$=0.008 and
log$L/L_{can}$=0.25, the average value suggested by the B07 Galactic
Cepheids.

%table 12
\linespread{0.7}
\begin{table*}\label{tab12}
\caption{Magellanic Cloud Cepheids with ISB-based distance. From left
to right, the columns give: name (1), period (2), ISB distance (3),
pulsation metal content log$Z$ and distance modulus $\mu_0$ at
log$L/L_{can}$=0 [(4)-(5)] and log$L/L_{can}$=0.25 [(6)-(7)]. The last
four columns give the final log$L/L_{can}$ ratio and metal content
derived by comparison with the ISB distance without [(8)-(9] or with
[(10)-(11)] the correction given by G05. In the last part of the
Table, we list the results for additional SMC Cepheids observed by
Laney \& Stobie (1994).}
\begin{center}
%\vspace{0.5truecm}
\begin{tabular}{lcccccccccc}
\hline \hline
\tiny{Name}&\tiny{log$P$}&\tiny{$\mu_{0,ISB}$}&\tiny{log$Z$}&\tiny{$\mu_{0,W}$}&\tiny{log$Z$}&\tiny{$\mu_{0,W}$}&\tiny{log$L/L_{can}$}&\tiny{log$Z$}&\tiny{log$L/L_{can}$}&\tiny{log$Z$}\\
           &             &\tiny{($\pm$)}   &   \tiny{$\pm$0.15}   &   \tiny{  $\pm$0.15   }   &   \tiny{  $\pm$0.15   }   &   \tiny{  $\pm$0.15   }   &   \tiny{  ($\pm$)             }   &   \tiny{  $\pm$0.15   }   &   \tiny{  ($\pm$) }   &   \tiny{  ($\pm$0.15  }   \\
          \tiny{1}
          &\tiny{2}&\tiny{3}&\tiny{4}&\tiny{5}&\tiny{6}&\tiny{7}&\tiny{8}&\tiny{9}&\tiny{10}&\tiny{11}\\
\hline\\
LMC\\\\
\tiny{HV12199}&\tiny{0.4215}&\tiny{18.336(0.094)}&\tiny{  $-$2.05}&\tiny{18.66}&\tiny{$-$2.09}&\tiny{18.44}&\tiny{+0.37(0.20)}&\tiny{ $-$2.12}&\tiny{+0.13}&\tiny{  $-$2.07   }   \\
\tiny{HV12203 }   &       \tiny{0.4704  }   &   \tiny{  18.481(0.092)   }   &   \tiny{  $-$1.87   }   &   \tiny{  18.73   }   &   \tiny{  $-$1.92   }   &   \tiny{  18.51   }   &   \tiny{  +0.29(0.20)        }   &   \tiny{  $-$1.93   }   &   \tiny{  +0.05    }   &   \tiny{  $-$1.88   }   \\
\tiny{HV12202 }   &       \tiny{0.4915  }   &   \tiny{  18.289(0.072)   }   &   \tiny{  $-$1.82   }   &   \tiny{  18.63   }   &   \tiny{  $-$1.87   }   &   \tiny{  18.41   }   &   \tiny{  +0.39(0.18)        }   &   \tiny{  $-$1.90   }   &   \tiny{  +0.16    }   &   \tiny{  $-$1.85   }   \\
\tiny{HV12197 }   &       \tiny{0.4975  }   &   \tiny{  18.165(0.058)   }   &   \tiny{  $-$2.02   }   &   \tiny{  18.70   }   &   \tiny{  $-$2.06   }   &   \tiny{  18.49   }   &   \tiny{  +0.62(0.18)        }   &   \tiny{  $-$2.14   }   &   \tiny{  +0.39    }   &   \tiny{  $-$2.09   }   \\
\tiny{HV12204 }   &       \tiny{0.5364  }   &   \tiny{  18.202(0.044)   }   &   \tiny{  $-$2.19   }   &   \tiny{  18.67   }   &   \tiny{  $-$2.24   }   &   \tiny{  18.46   }   &   \tiny{  +0.55(0.17)        }   &   \tiny{  $-$2.29   }   &   \tiny{  +0.32    }   &   \tiny{  $-$2.25   }   \\
\tiny{HV12198 }   &       \tiny{0.5469  }   &   \tiny{  18.314(0.028)   }   &   \tiny{  $-$1.98   }   &   \tiny{  18.73   }   &   \tiny{  $-$2.03   }   &   \tiny{  18.51   }   &   \tiny{  +0.48(0.17)        }   &   \tiny{  $-$2.07   }   &   \tiny{  +0.26    }   &   \tiny{  $-$2.03   }   \\
\tiny{HV12816 }   &       \tiny{0.9595  }   &   \tiny{  18.328(0.087)   }   &   \tiny{  $-$2.41   }   &   \tiny{  18.82   }   &   \tiny{  $-$2.46   }   &   \tiny{  18.60   }   &   \tiny{  +0.57(0.19)        }   &   \tiny{  $-$2.52   }   &   \tiny{  +0.43    }   &   \tiny{  $-$2.50   }   \\
\tiny{HV12815 }   &       \tiny{1.4169  }   &   \tiny{  18.296(0.028)   }   &   \tiny{  $-$1.86   }   &   \tiny{  18.58   }   &   \tiny{  $-$1.91   }   &   \tiny{  18.36   }   &   \tiny{  +0.33(0.17)        }   &   \tiny{  $-$1.93   }   &   \tiny{  +0.29    }   &   \tiny{  $-$1.92   }   \\
\tiny{HV879   }   &       \tiny{1.5662  }   &   \tiny{  18.532(0.040)   }   &   \tiny{  $-$1.97   }   &   \tiny{  18.76   }   &   \tiny{  $-$2.02   }   &   \tiny{  18.55   }   &   \tiny{  +0.27(0.17)        }   &   \tiny{  $-$2.02   }   &   \tiny{  +0.26    }   &   \tiny{  $-$2.02   }   \\
\tiny{HV2338  }   &       \tiny{1.6254  }   &   \tiny{  18.663(0.023)   }   &   \tiny{  $-$2.02   }   &   \tiny{  18.64   }   &   \tiny{  $-$2.07   }   &   \tiny{  18.43   }   &   \tiny{  $-$0.02(0.17)        }   &   \tiny{  $-$2.01   }   &   \tiny{  $-$0.02   }   &   \tiny{  $-$2.01   }   \\
\hline\\
SMC\\\\
\tiny{  HV1345  }   &       \tiny{1.1296  }   &   \tiny{  18.828(0.081)   }   &   \tiny{  $-$2.46   }   &   \tiny{  19.24   }   &   \tiny{  $-$2.51   }   &   \tiny{  19.02   }   &   \tiny{  +0.48(0.19)               }   &   \tiny{  $-$2.56   }   &   \tiny{  +0.38    }   &   \tiny{  $-$2.54   }   \\
\tiny{  HV1335  }   &       \tiny{1.1578  }   &   \tiny{  18.875(0.082)   }   &   \tiny{  $-$2.59   }   &   \tiny{  19.46   }   &   \tiny{  $-$2.64   }   &   \tiny{  19.25   }   &   \tiny{  +0.68(0.19)                }   &   \tiny{  $-$2.72   }   &   \tiny{ +0.59    }   &   \tiny{  $-$2.70   }   \\
\tiny{  HV1328  }   &       \tiny{1.1996  }   &   \tiny{  18.732(0.087)   }   &   \tiny{  $-$2.56   }   &   \tiny{  19.21   }   &   \tiny{  $-$2.61   }   &   \tiny{  19.00   }   &   \tiny{  +0.56(0.19)                }   &   \tiny{  $-$2.67   }   &   \tiny{  +0.48    }   &   \tiny{  $-$2.66   }   \\
\tiny{  HV1333  }   &       \tiny{1.2120  }   &   \tiny{  19.388(0.080)   }   &   \tiny{  $-$2.45   }   &   \tiny{  19.45   }   &   \tiny{  $-$2.50   }   &   \tiny{  19.23   }   &   \tiny{  +0.07(0.19)                }   &   \tiny{  $-$2.47   }   &   \tiny{  $-$0.01   }   &   \tiny{  $-$2.45   }   \\
\tiny{  HV822   }   &        \tiny{1.2238  }   &  \tiny{  19.091(0.081)   }   &   \tiny{  $-$2.24   }   &   \tiny{  19.27   }   &   \tiny{  $-$2.29   }   &   \tiny{  19.05   }   &   \tiny{  +0.20(0.19)                }   &   \tiny{  $-$2.28   }   &   \tiny{  +0.12    }   &   \tiny{  $-$2.27   }   \\
\hline
\tiny{  HV1365  }   &       \tiny{1.0940  }   &   \tiny{                  }   &   \tiny{  $-$2.39   }   &   \tiny{  19.50   }   &   \tiny{  $-$2.44   }   &   \tiny{  19.28   }   &   \tiny{                  }   &   \tiny{      }   &   \tiny{      }   &   \tiny{      }   \\
\tiny{  HV1954  }   &       \tiny{1.2230  }   &   \tiny{                  }   &   \tiny{  $-$2.70   }   &   \tiny{  18.89   }   &   \tiny{  $-$2.75   }   &   \tiny{  18.67   }   &   \tiny{                  }   &   \tiny{      }   &   \tiny{      }   &   \tiny{      }   \\
\tiny{  HV817   }   &       \tiny{1.2770  }   &   \tiny{                  }   &   \tiny{  $-$2.60   }   &   \tiny{  19.05   }   &   \tiny{  $-$2.65   }   &   \tiny{  18.83   }   &   \tiny{                  }   &   \tiny{      }   &   \tiny{      }   &   \tiny{      }   \\
\tiny{  HV11211 }   &       \tiny{1.3300  }   &   \tiny{                  }   &   \tiny{  $-$2.34   }   &   \tiny{  18.89   }   &   \tiny{  $-$2.40   }   &   \tiny{  18.67   }   &   \tiny{                  }   &   \tiny{      }   &   \tiny{      }   &   \tiny{      }   \\
\tiny{  HV2209  }   &       \tiny{1.3550  }   &   \tiny{                  }   &   \tiny{  $-$2.67   }   &   \tiny{  19.03   }   &   \tiny{  $-$2.72   }   &   \tiny{  18.81   }   &   \tiny{                  }   &   \tiny{      }   &   \tiny{      }   &   \tiny{      }   \\
\tiny{  HV847   }   &       \tiny{1.4330  }   &   \tiny{                  }   &   \tiny{  $-$2.35   }   &   \tiny{  19.23   }   &   \tiny{  $-$2.40   }   &   \tiny{  19.01   }   &   \tiny{                  }   &   \tiny{      }   &   \tiny{      }   &   \tiny{      }   \\
\tiny{  HV823   }   &       \tiny{1.5040  }   &   \tiny{                  }   &   \tiny{  $-$2.17   }   &   \tiny{  19.18   }   &   \tiny{  $-$2.22   }   &   \tiny{  18.96   }   &   \tiny{                  }   &   \tiny{      }   &   \tiny{      }   &   \tiny{      }   \\
\tiny{  HV865   }   &       \tiny{1.5230  }   &   \tiny{                  }   &   \tiny{  $-$2.74   }   &   \tiny{  18.96   }   &   \tiny{  $-$2.80   }   &   \tiny{  18.74   }   &   \tiny{                  }   &   \tiny{      }   &   \tiny{      }   &   \tiny{      }   \\
\tiny{  HV2064  }   &       \tiny{1.5270  }   &   \tiny{                  }   &   \tiny{  $-$2.32   }   &   \tiny{  19.32   }   &   \tiny{  $-$2.38   }   &   \tiny{  19.10   }   &   \tiny{                  }   &   \tiny{      }   &   \tiny{      }   &   \tiny{      }   \\
\tiny{  HV2195  }   &       \tiny{1.6210  }   &   \tiny{                  }   &   \tiny{  $-$2.50   }   &   \tiny{  19.12   }   &   \tiny{  $-$2.56   }   &   \tiny{  18.90   }   &   \tiny{                  }   &   \tiny{      }   &   \tiny{      }   &   \tiny{      }   \\
\tiny{  HV824   }   &       \tiny{1.8180  }   &   \tiny{                  }   &   \tiny{  $-$2.54   }   &   \tiny{  19.01   }   &   \tiny{  $-$2.60   }   &   \tiny{  18.80   }   &   \tiny{                  }   &   \tiny{      }   &   \tiny{      }   &   \tiny{      }   \\
\hline
\end{tabular}
\end{center}
\end{table*}
\linespread{1}

For the Magellanic Cepheids with ISB distances determined by G05, by
repeating the procedure adopted for the B07 and S04 Galactic Cepheids,
we derive the pulsation metal abundances and distances at
log$L/L_{can}$=0 and 0.25 listed in columns (4)-(5) and (6)-(7),
respectively, in Table 12. As a first result, we find that at constant
log$L/L_{can}$ ratio the pulsation distance to LMC Cepheids is
independent of the period, at variance with the ISB results listed in
column (3). On the other hand, taking at face value the ISB-based
distance to derive the log$L/L_{can}$ ratios listed in column (8), we
find log$L/L_{can}$=0.59$-$0.24log$P$ which disagrees with the
constant ratio already shown by the $P$-$WVI$ and $P$-$WJK$ relations
of the OGLE and P04's variables.  Interestingly enough, by comparing
the ISB distances with the pulsation distance moduli at
log$L/L_{can}$=0.25, which is the average value of the B07 Galactic
Cepheids, we derive
$$\Delta\mu_0(puls-ISB_{old})=0.29-0.20\log P\eqno(9)$$
\noindent which is surprisingly similar to the G05 relation given in
Eq. (3), but significantly different from the correspondent correction
to the S04 Galactic Cepheids given in Eq. (5). In other words, our
pulsation approach suggests that the G05 correction to ISB-based
distances holds for LMC Cepheids but would lead to deceptive results
(i.e., steeper $PL$ relations) if used for the (metal-richer) Galactic
Cepheids. Given the small number and the rather constant period of the
SMC Cepheids with ISB distance, no firm conclusion can be reached for
these metal-poorer variables.

Regarding the metal content inferred by the evolutionary $PW$
relations, the values at log$L/L_{can}$=0 listed in Table 12 give
log$Z=-$2.02$\pm$0.17 and $-$2.53$\pm$0.17 for LMC and SMC
Cepheids, respectively, with a fair decrease of 0.05 dex at
log$L/L_{can}$=0.25. These pulsation results agree with the widely
known average spectroscopic values [Fe/H]$\sim -$0.4 (LMC) and $-$0.7
(SMC).

\subsection{Comments on the universality of the $PL$ relations}

We have shown that the $HST$-based distances to Galactic Cepheids
support the predicted shallower slope of the $P$-$M_V$ and $P$-$M_I$ relations with
respect to the LMC relations, at odds with other empirical studies
supporting the universality of the $PL$ relations or, even worst,
pointing out that the Galactic relations are steeper than the LMC
ones.

In particular, the results of the Araucaria Project (see Pietrzynski
et al. 2007 and references therein) dealing with Cepheid observations
in Local Group galaxies suggest no significant change of the slope in
the oxygen abundance\footnote{[O/H]=log(O/H)-log(O/H)$_{\odot}$, with
log(O/H)$_{\odot}=-$3.1.} range from [O/H]=$-$0.4 to $\sim -$1.0,
i.e., {\it for lower metal abundances than $Z$=0.008.} This finding is
not inconsistent with the theoretical $P$-$M_V$ slopes which vary by
10\% from $Z$=0.02 to $Z$=0.008 but only 4\% from $Z$=0.008 to
$Z$=0.004, suggesting a sort of saturation in the metallicity effect
at very low metal abundances. As for the $P$-$M_I$ relation, the
predicted results already show no metallicity dependence below
$Z$=0.008. We plan to compute new pulsation models at $Z\le$ 0.002 to
verify this point.

Regarding the occurrence of steeper Galactic $PL$ relations, we have
shown that the S04 and G05 results are biased by the adopted
correction to ISB-based distances.  On the other hand, other papers,
e.g., Sandage et al. (2004) and Ngeow \& Kanbur (2004), adopt Cepheid
average distances from different available measurements which may
differ by up to 0.5 mag. Following a different approach to test the
universality of the $PL$ and $PW$ relations, let us adopt the observed
LMC relation $WVI$=15.88$-\mu_0(LMC)-$3.29log$P$ provided by the OGLE
data to derive the intrinsic distance modulus to each Galactic Cepheid
from its measured $WVI$ function. Once the apparent $V$ and $I$
magnitudes are corrected for reddening, this method provides the
absolute Galactic $PM_V$ and $PM_I$ relations for each given LMC
distance.

%\fig14
\begin{figure}
\includegraphics[width=8cm]{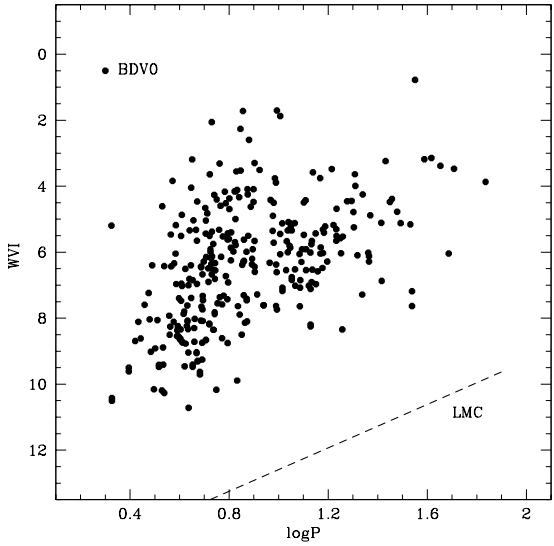}
\caption{\small{Observed $WVI$ functions versus period for fundamental Galactic
Cepheids with measured $V$ and $I$ magnitudes compiled by Berdnikov,
Dambis \& Vozyakova (2000 [BDV0]). The dashed line is the observed
LMC Cepheid relation.}} \label{7587fig14}
\end{figure}

%\fig15
\begin{figure}
\includegraphics[width=8cm]{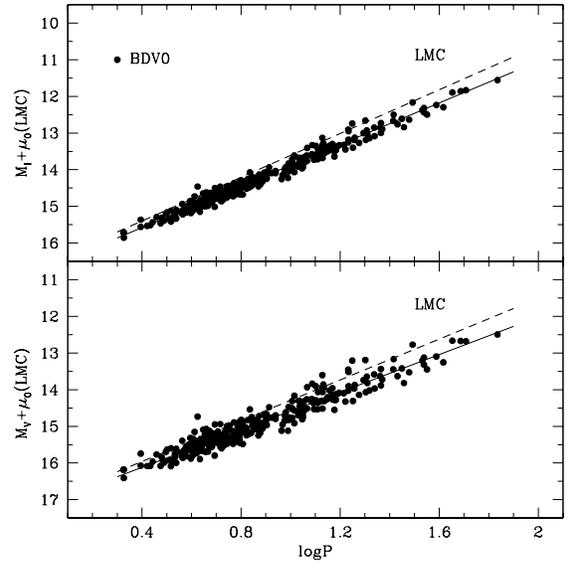}
\caption{\small{$M_V$ and $M_I$ magnitudes versus period for the
Galactic Cepheids in Fig. 14. The solid lines are drawn by a linear
regression through the points, while the dashed lines are the LMC
Cepheid relation.}} \label{7587fig15}
\end{figure}

Figure 14 shows all the fundamental Galactic Cepheids listed by
Berdnikov, Dambis \& Vozyakova (2000) in comparison with the
observed $P$-$WVI$ relation of LMC Cepheids. 
Adopting the $E(B-V)$ reddening listed by Fernie et al. (1995), we
show in Fig. 15 that the resulting relations (solid lines) are
$$M_V=17.14(\pm0.17)-\mu_0(LMC)-2.56(\pm0.17)\log P\eqno(10)$$
$$M_I=16.71(\pm0.12)-\mu_0(LMC)-2.83(\pm0.10)\log P\eqno(11)$$
\noindent 
This result clearly demonstrates that assuming that
the LMC Wesenheit relations are universally valid lead to absolute
Galactic $PM_V$ and $PM_I$ relations whose slope are significantly
different from the LMC ones. In particular, the Galactic relations 
are not only {\it
shallower  than the LMC $PM_V$ and $PM_I$ relations (dashed lines)
given in Table 11, but even fainter (at log$P\ge$ 0.5), whichever
is the LMC distance.}}

\section{Conclusions}
{\bf We have computed new sets of nonlinear fundamental pulsation models for
Classical Cepheids assuming four different chemical compositions
($Z$=0.004, 0.008, 0.01 and 0.02)
and adopting an increased value
($l/H_p$=1.7-1.8) of the mixing length parameter used in the
hydrodynamical code to close the system of nonlinear equations. By
combining the results of these calculations with our previously
computed models at $l/H_p$=1.5 we find that}:
\begin{enumerate}
\item {\bf increasing the $l/H_p$ value the Cepheid instability strip gets narrower and
the effect gets more important at the higher metal contents ($Z\ge$ 0.01). However,
the synthetic Period-Magnitude relations based on the combination of pulsation
results with updated evolutionary models are slightly affected
by the adopted $l/H_p$ value and confirm  the already shown dependence
of the slope, the zero point and the intrinsic dispersion on the
wavelength and the adopted metallicity;}
\item {\bf by using all the fundamental models computed with different choices of the
metal and helium abundance, we determine mass-dependent Period-Wesenheit
relations which are useful to estimate the actual mass of Cepheids
of known distance and metal content};
\item {\bf in order ot avoid any assumption on the Cepheid Mass-Luminosity relation, we
introduce the ratio log$L/L_{can}$, where $L_{can}$ is the luminosity inferred by
canonical (no mass-loss, no convective core overshooting) evolutionary computations,
and we derive evolutionary Period-Wesenheit relations which provide
the distance to Cepheids with known metal content, once the $L/L_{can}$ ratio is
adopted. As discussed in the text, the release of the canonical evolutionary frame
yields positive log$L/L_{can}$ values};
\item {\bf we show that the effects of
$\l/H_p$ (at fixed $Z$ and
$Y$) and $Y$ (at fixed $\l/H_p$ and $Z$) on the evolutionary Period-Wesenheit
relations are quite negligible. In the meanwhile, we show that
the predicted $P$-$WBV$ relation has an opposite metallicity
dependence with respect  to $P$-$WVI$ and $P$-$WVK$, suggesting a quite
plain method to derive the distance and metallicity of observed variables,
independently of reddening}.
\end{enumerate}

{\bf By comparison of the theoretical relations with Galactic and LMC Cepheids, we derive the
following results}:
\begin{enumerate}
\item {\bf the pulsation distances inferred from the predicted evolutionary
$PW$ relations
are consistent with the recent $HST$ astrometric measurements given by Benedict et al.
(2007) for a sample of
Galactic Cepheids and suggest an average value of log$L/L_{can}$=0.25$\pm$0.12 or a
mild period-dependence as log$L/L_{can}=0.41-$0.19log$P$};
\item {\bf on the contrary, the pulsation distances disagree with those
determined by Storm et al. (2004) for Galactic Cepheids, as based on the
Infrared Surface Brightness technique, unless
negative (unrealistic) log$L/L_{can}$ ratios and
log$L/L_{can}=0.85-$0.55log$P$ are adopted};
\item {\bf adopting for
these variables log$L/L_{can}$=0.25,
the average value inferred from the Benedict et al.
(2007) Cepheids, we determine a
period-dependent correction to the
ISB-based distances as given by
$$\Delta\mu_0(puls-ISB_{old})=0.50-0.45\log P,$$
\noindent
which is significantly different from the correction
$$\Delta\mu_0(ISB_{new}-ISB_{old})=0.29-0.18\log P$$
determined by Gieren et al. (2005) to remove the puzzling dependence
on the period of the ISB distances to LMC Cepheids. It is
of interest to note that applying our pulsation approach to these LMC Cepheids, we
actually get
$$\Delta\mu_0(puls-ISB_{old})=0.29-0.20\log P,$$
\noindent suggesting an additional metallicity effect on the
period-dependent correction to ISB distances;} \item {\bf the
reliability of the pulsation method to infer the Cepheid
metallicity is verified by comparison with available spectroscopic
data for Galactic and LMC Cepheids}; \item {\bf the slopes of the
observed Period-Magnitude relations of the hundreds of LMC
Cepheids observed by Udalski et al. (1999) and Persson et al.
(2004) are in close agreement with the those predicted at
$Z$=0.008. A similar agreement is found for the Period-Wesenheit
relations, which also exclude a significant variation of the
$L/L_{can}$ ratio among the LMC Cepheids}; \item {\bf the
comparison of predicted with observed $P$-$WVI$ and $P$-$WJK$
relations gives $\mu_0$(LMC)=18.73$\pm$0.09 mag at
log$L/L_{can}$=0 and 18.52$\pm$0.11 mag at log$L/L_{can}$=0.25,
the average value suggested by the B07 Galactic Cepheids}; \item
{\bf the predicted steepening of the optical Period-Magnitude
relations when passing from metal-rich to metal-poor variables is
supported by the $HST$-based distances to Galactic Cepheids, as
well as by observational evidences based on the straight
application to Galactic Cepheids of the {\it empirical} LMC
$P$-$WVI$ relation. Moreover, we show that at log$P\ge$ 0.5 the
Galactic $P$-$M_V$ and $P$-$M_I$ relations are fainter than the
LMC ones in agreement with the theoretical predictions.}
\end{enumerate}

\begin{acknowledgements}
We thank the anonymous referee for several valuable comments and suggestions.
It is a pleasure to acknowledge useful conversations with Michele Cignoni and
Giuseppe Bono.
Financial support for this study was provided by MIUR, under the
scientific project ``Continuity and Discontinuity in the Milky Way
Formation'' (PI: Raffaele Gratton).
\end{acknowledgements}

\pagebreak
%\begin{references}

%\end{references}

\end{document}